\documentclass[10pt,journal,final,twocolumn]{IEEEtran}

\usepackage{amsmath}
\usepackage{amssymb}
\usepackage{graphicx}
\usepackage{subfigure}
\usepackage{makecell}
\usepackage{multirow}
\usepackage{cite}
\usepackage{algorithm}
\usepackage{algorithmicx}
\usepackage{algpseudocode}
\usepackage{cases}
\usepackage{stfloats}
\usepackage{enumerate}
\usepackage{url}
\usepackage{booktabs}
\usepackage{diagbox}
\usepackage{bbding}
\usepackage{color}
\usepackage{xcolor}
\usepackage{framed}

\allowdisplaybreaks[4]

\makeatletter
\renewcommand{\maketag@@@}[1]{\hbox{\m@th\normalsize\normalfont#1}}
\makeatother



\renewcommand{\color}[1]{}

\begin{document}	
    \title{
    Synchronization, Identification, and Signal Detection for Underwater Photon-Counting Communications With Input-Dependent Shot Noise
    }
	\author{Fanghua~Li,
      Xiaolin~Zhou,~\IEEEmembership{Senior Member,~IEEE},
      Yongkang~Chen,~\IEEEmembership{Member,~IEEE},
      Wei~Ni,~\IEEEmembership{Fellow,~IEEE},\\
	 Xin~Wang,~\IEEEmembership{Fellow,~IEEE},
	 Dusit Niyato,~\IEEEmembership{Fellow,~IEEE},		
	 and~Ekram Hossain,~\IEEEmembership{Fellow,~IEEE}
		\thanks{
        This work was supported by the National Natural Science Foundation of China under Grant 62571137, 62231010 and 62501381, and in part by the Natural Science Foundation of Shanghai under Grant 24ZR1407100.
        \textit{(Corresponding author: X. Zhou.)}}
		\thanks{F. Li, X. Zhou,
			and X. Wang are with Key Laboratory for Information Science of Electromagnetic Waves, School of Information Science and Technology, Fudan University, Shanghai 200433, China (Email: 21110720109@m.fudan.edu.cn; zhouxiaolin@fudan.edu.cn; xwang11@fudan.edu.cn).}
        \thanks{Y. Chen is with the School of Information Engineering, Shanghai Maritime University, Shanghai 201306, China (Email: ykchen@shmtu.edu.cn).}
		\thanks{W. Ni is with the Commonwealth Scientific and Industrial Research Organization, Sydney, NSW 2122, Australia (Email: wei.ni@data61.csiro.au).}
\thanks{D. Niyato is with the College of Computing and Data Science, Nanyang Technological University, Singapore 639798 (Email: dniyato@ntu.edu.sg).}        
\thanks{E. Hossain is with the Department of Electrical and Computer Engineering, University of Manitoba, Winnipeg, MB R3T 2N2, Canada (Email: ekram.hossain@umanitoba.ca).}
	}

	\maketitle
 
    \begin{abstract}
        Photon counting (PhC) is an effective detection technology for underwater optical wireless communication (OWC) systems. The presence of signal-dependent Poisson shot noise and asynchronous multi-user interference (MUI) complicates the processing of received data signals, hindering the effective signal detection of PhC OWC systems. This paper proposes a novel iterative signal detection method in grant-free, multi-user, underwater PhC OWC systems with signal-dependent Poisson shot noise. We first introduce a new synchronization algorithm with a unique frame structure design. 
        The algorithm performs active user identification and transmission delay estimation. Specifically, the estimation is performed first on a user group basis and then at the individual user level with reduced complexity and latency.
        We also develop a nonlinear iterative multi-user detection (MUD) algorithm that utilizes a detection window for each user to identify interfering symbols and estimate MUI on a slot-by-slot basis, followed by maximum \textit{a-posteriori} probability detection of user signals.
        Simulations demonstrate that our scheme achieves bit error rates comparable to scenarios with transmission delays known and signal detection perfectly synchronized.    
    \end{abstract}


    \section{Introduction}	
	\label{Sec:Intro}

\IEEEPARstart{U}{nderwater} optical wireless communication (OWC) has gained significant attention due to its high bandwidth~\cite{COMST:2021UnderwaterSurvey,ICC:2021Underwater}, high transmission rate~\cite{COMST:2021UnderwaterSurvey}, and low latency~\cite{ICC:2021Underwater}. Moreover, OWC offers an eco-friendly alternative by minimizing the use of sonar and hydroacoustic equipment, which could disturb marine animals sensitive to acoustic signals~\cite{COMST:2021UnderwaterSurvey}. Experimental studies in the Baltic Sea~\cite{UComms:scholz2018} have demonstrated the feasibility of underwater OWC in practice. In seawater, elevated background radiation and turbulence increase absorption and scattering~\cite{COMST:2021UnderwaterSurvey}, which can significantly reduce received power~\cite{CL:2017wang} and complicate signal detection.
In low-power underwater OWC scenarios, signal-dependent shot noise becomes the primary performance-limiting factor, deviating markedly from the traditional additive white Gaussian noise (AWGN) model commonly used in radio frequency (RF) systems~\cite{TCOMM:2005wilson}. Photon counting (PhC), with its superior detection capabilities for low-intensity optical pulses, emerges as an effective and reliable signal detection technique for underwater OWC system receivers~\cite{PJ:huang2021}.  

In marine development, there is a growing need for timely responses to tasks, e.g., oil and gas exploration, pipeline monitoring, and disaster warning~\cite{Sensors:2021DAMAC}. To address these needs, various underwater schemes have been proposed~\cite{Sensors:2021DAMAC,JMSE:2023asynchronous}. Unlike traditional access methods, grant-free schemes allow users to transmit signals asynchronously without waiting for a scheduling grant from a base station (BS). While simplifying communication protocols and device designs, grant-free schemes lead to overlapping signals received at the BS~\cite{ref6}. 

\begin{figure}[!t]
        \centering
        \includegraphics[width=0.33\textwidth]{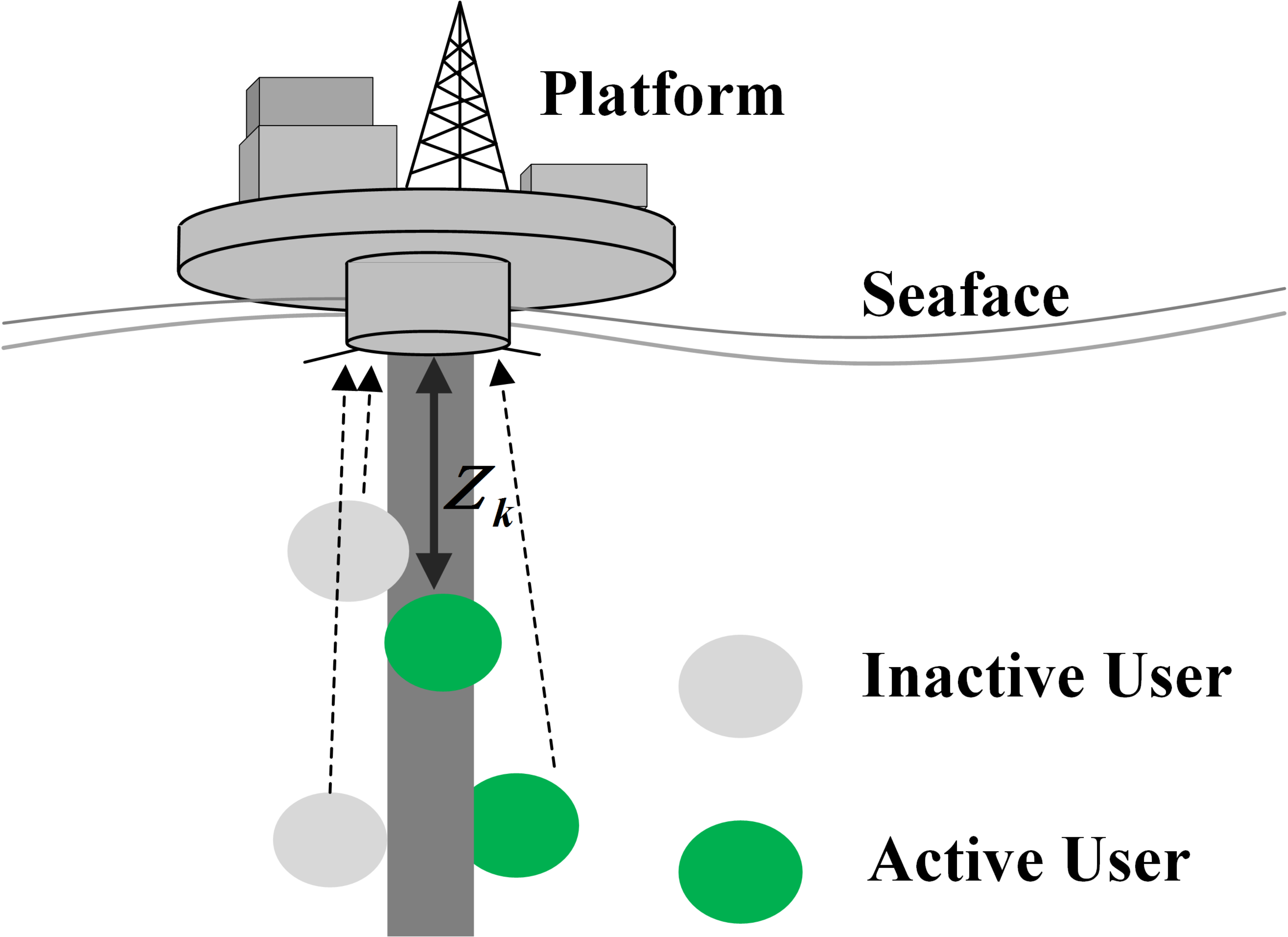}
        \caption {Illustration of grant-free, multi-user, underwater PhC OWC systems.}
        \label{fig:Scenario}
    \end{figure}

{\color{blue}
    \begin{table*}    
\centering
\caption{\color{blue}Review of Existing Techniques}
\label{TableContri}
		\begin{tabular}{|l|c|c|c|c|c|c|c|}  
			\hline
			\quad  
			& [18]
			& [19]
			& [14]
			&This paper\\
			\hline

            \makecell{\color{blue}Iterative detection}
			& $\surd$ 
			& $\surd$ 
			& 
			& $\surd$
			\\
			\hline
			
			\makecell{\color{blue}Delay estimation}
			&  
			&  
			& $\surd$
			& $\surd$
			\\
			\hline

            \makecell{\color{blue}Iterative detection complexity}
            & ${T_{{\rm{out}}}}{L_{\rm{s}}}{K^2}$ 
            & ${T_{{\rm{out}}}}{L_{\rm{s}}}{K^2}$ 
            & N/A
            & ${T_{{\rm{out}}}}{L_{\rm{s}}}K_{\rm{a}}^2$ \\
            \hline

            \makecell{\color{blue}Delay estimation complexity}
            & N/A 
            & N/A 
            & \makecell{$Q{(B{N_{\rm{c}}})^2} + {(B{N_{\rm{c}}})^3} $ \\ $ + K{N_{\rm{c}}}(B{N_{\rm{c}}})$  
            }
            & \makecell{$ K L_{\rm{p}} L_{\rm{s}} + T_{\rm{in}} \big[ {\max_{\forall m}}(K_{\mathrm{a},m} K_m) $ \\ 
            $ + {\max_{\forall m}}(M K_m L_{\rm{q}}) \big]$} \\
            \hline
            
			
			\makecell{\color{blue}Convergence (iterations)}
			& 5 
			& 5 
			& N/A
			& 5
			\\
			\hline
			
			\makecell{\color{blue}BER ($E_{\rm{b}} = -166$ dBJ, Poisson model)} 
			& $ {10^{ - 3}}$ 
			& $ {10^{ - 3}}$ 
			& $ {10^{ - 2}}$
			& ${10^{ - 6}}$
			\\
			\hline
			
		\end{tabular}	
    \smallskip \\
    \footnotesize{\color{blue}
    $B$: snapshot length, $Q = L_{\mathrm{c}}/B$: number of snapshots. The complexity of the proposed method is analyzed in Sections~III-D and~IV-D. 
    }    
	\end{table*}
    }

    \subsection{Related Work}
    \label{Related Work}
    Underwater PhC OWC systems pose significant challenges due to the quantum effects~\cite{PR:1967Fray} and signal-dependent Poisson shot noise~\cite{TIT:2021ahmadypour},~\cite{TCOM:ChenYK2023}, which differ from the widely studied signal-independent AWGN models. 
    When users are allowed to transmit in a grant-free manner, 
    identifying transmitting users and then performing multi-user detection (MUD) for their signals are critical. The study in~\cite{WN:2023qiu} investigated an underwater Internet-of-Things (IoT) acoustic communication system, where user activity is estimated by designing pilot signals based on the classical AWGN model. The direct detection approach under the AWGN assumption in acoustic systems differs fundamentally from the PhC OWC systems. The authors in~\cite{PJ:2024experimental} proposed a subspace-based method to estimate user delays in asynchronous (grant-free) multi-user underwater OWC systems under the AWGN model. This blind estimation method is computationally intensive, making it less suitable for low-power underwater receivers.

    Signal demodulation in PhC OWC systems is challenging due to the presence of Poisson shot noise. Unlike the signal-independent noise in the traditional AWGN models, receiver noise in discrete-time Poisson (DTP) channels depends on the received signal, with the noise variance increasing proportionally to the signal level~\cite{TCOM:ChenYK2023sensorNetwork}. This signal-dependent nature complicates analysis, as no tractable analytical expression exists for the signal-to-interference-plus-noise ratio (SINR) in a DTP channel~\cite{TCOM:ChenYK2023sensorNetwork}. Classic SINR-based MUD methods, such as smooth demodulation, energy detection, and matched filtering~\cite{CL:2006new}, are ineffective. The message passing algorithm (MPA) based on maximum likelihood (ML) detection offers near-optimal MUD performance but suffers from exponential complexity with an increasing number of users~\cite{CL:2016yang}. 
    
    Several low-complexity iterative interference cancellation techniques have been proposed in the literature~\cite{OE:zhouXL2012},~\cite{ICICN:2019li}. These methods typically spread each user’s bit sequence using a simple repetition code to generate a signal sequence, which is then interleaved using a user-specific interleaver. The receiver first employs an elementary signal estimator (ESE) for all users, followed by a dedicated \textit{a-posteriori} probability detector (APP-DEC) for each user~\cite{TC:1996Be}. 
    In~\cite{JOE:2011aliesawi}, iterative MUD techniques were explored for underwater wireless communication systems to mitigate user interference.
    {\color{blue} As shown in Table~\ref{TableContri}, the proposed method achieves lower computational complexity (due to grouping, $K_{\mathrm{a}} \ll K$) and significantly improved BER performance, compared to the existing schemes under identical Poisson shot noise conditions.}

    In conventional grant-based NOMA systems, uplink access requires request, grant, and data transmission, leading to queuing, scheduling, and channel access delays. Collisions require additional retransmissions, further increasing delay and reducing the success probability of data transmission~\cite{TCOM:2022zhou}. In contrast, grant-free schemes eliminate the request-grant handshake after synchronization, avoiding the additional transmission delay after communication establishment.
    Moreover, grant-free transmissions lack the timing control signals present in grant-based access, causing user signals to arrive at BS with varying delays and resulting in asynchronous multi-user interference (MUI). Each user experiences MUI not only from overlapping signals of other users but also from its own signals. The study in~\cite{ref6} analyzed asynchronous interference elimination for receivers and derived interference signals in an AWGN context. However, these studies have not addressed receiver design for asynchronous, grant-free, multi-user PhC OWC systems limited by DTP noise.

    \begin{figure*}[!t]
        \centering
        \includegraphics[width=0.9\textwidth]{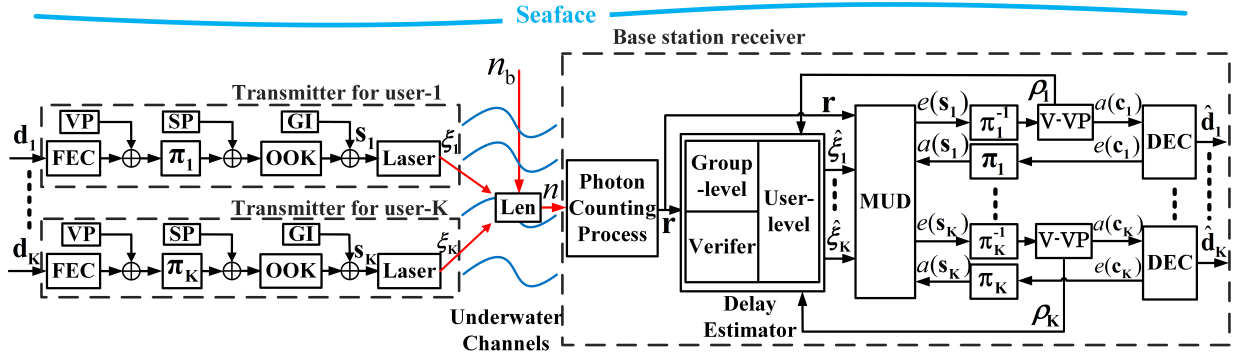}
        \caption {The system architecture and modules of the considered grant-free, multi-user, underwater PhC OWC systems.}
        \label{fig:SystemModel}
    \end{figure*}

    \subsection{Contribution}
    \label{Contribution}    
    This paper proposes a low-complexity iterative detection method for grant-free, multi-user, underwater, PhC OWC systems
    under signal-dependent Poisson shot noise and strong background radiation. By identifying active users and estimating their delays, the proposed method can reduce the bit error rate (BER) of the systems significantly.    
    The contributions of this paper are summarized as follows.

    \begin{itemize}
        \item We design a new grant-free, multi-user, underwater PhC OWC system under the signal-dependent Poisson shot noise, where users can transmit asynchronously in a grant-free manner. The BS can effectively synchronize and iteratively detect the asynchronous
        signals.
        
        \item 
        We develop a new synchronization algorithm, incorporating a new frame structure to detect user activity and estimate delays.
        A verification step addresses delay estimation errors resulting from Poisson shot noise. 
        An efficient grouping strategy reduces algorithmic complexity, significantly shortening synchronization time. 
        
        \item  
        We develop an iterative MUD algorithm with delay adjustment to suppress MUI and Poisson shot noise, where a detection window is employed to accurately identify interfering symbols. We design a slot-by-slot iterative log-likelihood ratio (LLR) detection algorithm, based on the maximum \textit{a-posteriori} probability principle, to recover asynchronous signals. 
  
    \end{itemize}    
    
    Simulations demonstrate effectiveness of the proposed approach under grant-free, asynchronous transmissions and Poisson shot noise. 
    Compared to direct detection methods, such as classic MMSE or Gaussian approximation, which struggle in asynchronous scenarios without delay estimation and calibration, the proposed iterative MUD algorithm offers substantial advantages by iteratively refining the detection process in support of asynchronous transmission. The algorithm achieves fast convergence within up to five iterations, demonstrating its practical effectiveness.

\begin{table} [!t]
    \caption{Notation and Definitions}
    \label{table:Notation}
    \begin{center}
    \begin{tabular}{ >{\centering\arraybackslash}m{1.6cm} >{\raggedright\arraybackslash}m{6.4cm}}
        \hline
        Notation & Description \\
        \hline
        ${\alpha}_{k}$, ${\alpha}_{k}(j)$, $\hat{\alpha}_{k}(j)$ & Activity of user $k$, its $j$-th window activity, and accuracy-adjusted estimated activity in window $j$ \\
        $a(\mathbf{c}_k)$,$e(\mathbf{c}_k)$, $a(\mathbf{s}_k)$, $e(\mathbf{s}_k)$ & The \textit{a-priori} LLR and extrinsic LLR for user $k$, respectively, in DEC and MUD\\        
        $C$, $L_k$, $G_k$, $I_k$, $Z_k$ & Attenuation coefficient, path loss, overall channel gain, fading coefficient and user distance for user~$k$ \\
        $\mathbf{d}_k$, $d_{k}(i)$ & Data sequence and its $i$-th bit for user $k$ \\
        $\hat{d}_{k,j}$ & Recovered bit for user $k$ in frame $j$ \\
        $E_{\mathrm{b}}$, $E_{\mathrm{nb}}$, $N_{\mathrm{c}}$ $\;$ & Energy per bit received and background radiation in each slot, and length of the spread spectrum sequence \\
        $f$ & Optical center frequency \\
        $j$ & Frame/detection window index \\
        $K$, $k$ & Number and index of users \\
        $K_a$,$K_m$,$K_{a,m}$ $\;$ & Number of active users, total number of users in group $m$ and number of active users in group $m$ \\
        $L_{\mathrm{b}}$, $L_{\mathrm{q}}$, $L_{\mathrm{p}}$, $L_{\mathrm{s}}$ & Lengths of data, verification pilot, synchronization pilot and transmitted packet, respectively\\
        $L_{\mathrm{A}}$, $L_{\mathrm{E}}$ & The \textit{a-priori} and extrinsic LLRs of SISO \\
        $I_{\mathrm{A}}$, $I_{\mathrm{E}}$ & The MI between the transmitted symbol $s$ and the \textit{a-priori} and extrinsic LLRs for SISO \\
        ${\cal I}({\xi _k}(j))$ & The Fisher information for ${\xi _k}(j)$ \\
        ${\color{blue}\hat{G}_k}$ & {\color{blue}The estimated CSI for user $k$} \\
        $\mu_{k}(j)$,$u_{k}(j)$, $\upsilon_{k}(j)$, $\nu_{k}(j)$ & Delay mean, prior mean, likelihood mean and posterior mean for user $k$ in window $j$ \\
        $M$, $M_k$ & Number of user groups and group index for user $k$ \\
        $n(l)$, $n_k$& Expected photons in slot $l$ and the expected photon \\
                     & count for user $k$\\
        $n_{\mathrm{s}}$, $n_{\mathrm{b}}$ & Expected photons of desired and background radiation \\
        $\eta$& Quantum efficiency\\
        $P_{\mathrm{s}}$ & Transmit power of the laser \\
        ${\rho}_{k}(j)$, $\hat{\rho}_{k}(j)$ & Delay verification factor and estimated delay \\
        & verification factor for user $k$ in window $j$ \\
        $\mathbf{q}$, $q(l)$ & Verification pilot and its $l$-th slot \\
        $\mathbf{p}_m$, $p_{m}(l)$ & Synchronization pilot and its $l$-th slot for group $m$ \\
        $R_{k}(j)$, $R_{m,j}(n)$ & Correlation between received and transmitted VP for user $k$ in frame $j$, and correlation between received and transmitted SP in the $n$-th slot of group $m$ and \\
        & frame $j$ \\ 
        $r(l)$, ${r}_{j}(l)$ & Photo-electron count at slot $l$ and in frame $j$ \\
        $r_{k,j}(l)$ & Photo-electron count at slot $l$, in frame $j$ for user $k$ \\
        $\sigma_x^2$, $\sigma_x$ & Variance and intensity of fading coefficient $I_k$ \\
        $\sigma_{k}^2(j)$,$\delta_{k}^2(j)$, $\phi_{k}^2(j)$,$\zeta_{k}^2(j)$,& Delay variance, prior variance, likelihood variance and posterior variance for user \(k\) in window \(j\) \\    
        $\varphi_{k}^2$ & Variance of the delay of user $k$ \\
        $\mathbf{s}_k$, $s_{k}(l)$ & Transmitted sequence and its $l$-th slot for user $k$ \\
        $\tau$ & Slot duration \\
        $\Tilde{\mathbf{q}}$, $\Tilde{q}(l)$ & Extracted verification pilot and its $l$-th slot \\
        $\Tilde{r}_{j}(l)$, & Expected desired signal for user~$k$ in frame~$j$\\
        ${\xi}_{k}$, ${\xi}_{k}(j)$ & Delay and window-specific delay  for user $k$ \\
        ${\Xi}_{m}(j)$, $\Tilde{\xi}_{k}(j)$, $\hat{\xi}_{k}(j)$ & Estimated delay set, estimated delay and adjusted estimated delay for user~$k$ in window $j$, respectively \\
        $\varepsilon_{\mathrm{p}}$, $\varepsilon_{\mathrm{q}}$, ${\color{blue}{\varepsilon_k}}$ & Delay estimation threshold, delay verification threshold, {\color{blue}and CSI estimation error for user~$k$} \\   
        \hline
    \end{tabular}
    \end{center}
\end{table}
    The rest of this paper is organized as follows. Section~\ref{Sec:SysModel} introduces underwater OWC systems. In Section \ref{Sec:randomaccess}, we design the synchronization algorithms for group-level and user-level delay estimation, and delay verification, and analyze the algorithm complexity. In Section \ref{sec:Iterative detection}, we design an iterative MUD receiver for asynchronous signals. Simulations are provided in Section \ref{sec:rsults}. Conclusions are drawn in Section~\ref{sec:conclusion}. Table~\ref{table:Notation} summarizes the notation used.
    
\section{System Model}
\label{Sec:SysModel}
  
As illustrated in Fig.~\ref{fig:Scenario}, the considered grant-free, multi-user, underwater PhC-OWC system comprises $K$ users and a platform that serves as the BS. Each user has the same activation probability, 
as typically assumed~\cite{TVT:zhaobo2020}. Compared to atmospheric transmission, the quality and composition of water significantly affect light attenuation, hence degrading the received signal quality substantially~\cite{TCOM:ChenYK2023sensorNetwork}. This primarily impacts channel modeling, as will be delineated in Section~\ref{sec:channel}. 
The users follow a grant-free protocol, and transmit data in an uncoordinated manner. Their transmissions can be asynchronous and overlap.
The architecture of the system is depicted in Fig.~\ref{fig:SystemModel}.

\subsection{Transmitter}
\label{sec:transmitter}

As illustrated in Fig.~\ref{fig:SystemModel}, the bit sequence of user $k$ is denoted by $\mathbf{d}_k = \{{d_{k}(i), i = 1, \ldots, L_{\mathrm{b}}}\}$, where $L_{\mathrm{b}}$ is the frame length (in bits). The forward error correction (FEC) encoder generates the coded sequence. A \textit{verification pilot (VP)} is introduced to verify the activity of user $k$. We assume all users use the same VP. The VP, $\mathbf{q} = \{q_l, \, l = 1, \ldots, L_{\mathrm{q}}\}$, can be a random bit sequence shared by all users, followed by the coded bit sequence.
As illustrated in Fig.~\ref{fig:FrameStructure}, green and black represent the VP and coded bit sequences, respectively. 
The coded bit sequence and the VP are combined and passed through an interleaver, $\Pi_k$, to produce the interleaved sequence.

To facilitate the detection of asynchronous transmissions at the receiver, a \textit{synchronization pilot (SP)} is combined with the interleaved sequence. The $K$ users are divided into $M$ groups. 
Users in the $m$-th group share the same SP, denoted as ${\mathbf{p}_{m}}$. The SP, $\mathbf{p}_m = \{p_{m}(l), \, l = 1, \ldots, L_{\mathrm{p}}\}$, is marked in orange in Fig.~\ref{fig:FrameStructure}.
Let $\mathcal{U}_m$ collect the indexes of the users in the $m$-th group sorted in ascending order. $\mathcal{U}_m(k')$ indicates the $k'$-th user in the $m$-th group. 

On-off keying (OOK) modulation is applied, resulting in modulated symbol sequences transmitted in slots (one symbol per slot). A guard interval (GI) is added between frames to prevent inter-frame interference (IFI). 
The transmitted symbol sequence $\mathbf{s}_k = \{{s_{k}(l), \,l = 1, \ldots, L_{\mathrm{s}}}\}$ drives the laser, emitting an optical signal into underwater channels. The transmit power is $P_{\mathrm{s}}$. The expected number of emitted photons is $n_{\mathrm{s}} = \eta \frac{P_{\mathrm{s}} \tau}{hf}$, where $\eta$, $\tau$, $h$, and $f$ are the efficiency of the photon-conversion process, the slot duration, Planck’s constant, and the optical signal frequency, respectively. 
{\color{blue}According to~\cite{TCOM:2019Zou}, the dead-time effect refers to the inability of imperfect photon-counting detectors to register photons arriving within a short recovery time. Nevertheless, multi-pixel photon counters (MPPCs) greatly mitigate this limitation by employing a parallel array of microcells. Commercial devices, such as the Hamamatsu S13360 series, contain thousands to tens of thousands of microcells~\cite{OC:2019underwater},~\cite{Hamamatsu:2025}, which is significantly higher than the maximum photon count per symbol in our system (in the order of a few hundred). Thus, the dead-time effect is effectively mitigated.}

\subsection{Underwater Channel}
\label{sec:channel}
 The underwater environment is relatively stable compared to the surface environment, with user-specific transmission paths resulting in varying channel gains between each user and the BS~\cite{GLOBECOM2011channel}. The fading coefficient $I_k$ between user $k$ and the BS follows a log-normal marine model with mean 1 and variance $\sigma_x^2$. Its probability density function (PDF) is given by~\cite{TCOM:ChenYK2023sensorNetwork}
\begin{equation}
\label{deqn_ex1a1}
\Pr(I_k = i) = \frac{1}{\sqrt{2\pi} i\sigma_x} \exp\left[-\frac{1}{2\sigma_x^2} \left( \ln i + \frac{\sigma_x^2}{2} \right)^2 \right], \,\forall i > 0,
\end{equation}
where $\sigma_x$ denotes the turbulence intensity.

The path loss $L_k$ between user $k$ and the BS follows the Beer-Lambert law and experiences exponential attenuation with distance $Z_k$~\cite{OJVT:2025Zhang}:
\begin{equation}
L_k = \exp(-C Z_k),
\end{equation}
where $C$ is the attenuation coefficient, which depends on the water type. Assuming a clear ocean environment, the attenuation coefficient is set to $C = 0.15$ m$^{-1}$~\cite{BookConvexOpt}.

The channel gain from the BS to user $k$ is $G_k = I_k L_k$. For efficient decoding, we assume that the receiver has full knowledge of the channel state information (CSI).
    
 \subsection{Receiver}
    \label{sec:receive}
    
    At the receiver, a photon-detector (PD) is used to detect 
    optical signals. The PD converts incoming photons into photo-electrons and generates electrical pulses to trigger a built-in counter~\cite{TCOM:ChenYK2023}. Due to quantum effects~\cite{PR:1967Fray}, the PhC 
    follows a Poisson process~\cite{OE:zhouXL2012}, with the probability given by~\cite{TCOMM:2005wilson}:  
\begin{equation}
\Pr\left( r(l) \mid s_{k}(l) \right) = \frac{\left( n(l) \right)^{r(l)}}{r(l)!} \exp\left(-n(l)\right),  
\label{eq:poissionPDF}
\end{equation}  
where \( r(l) \) is the photo-electron count in the \( l \)-th slot, and \( n(l) \) is the expected number of received photons.  

The expected photon count, \( n(l) \), is given by 
\begin{equation}
n(l) = n_{\mathrm{b}} + \sum_{k=1}^K \alpha_k n_k s_{k}(l),
\end{equation} 
where \( n_{\mathrm{b}} \) is the expected photon count from background radiation, \( \alpha_k\) denotes the activity of user \( k \), \( n_k \) is the expected photon count from user \( k \)’s signal, and \( s_{k}(l) \) is the \( l \)-th symbol transmitted by user \( k \) (in the $l$-th slot).

Background radiation accounts for environmental factors and the PD's dark current. The expected photon count from user \( k \)’s signal, \( n_k \), is given by 
\begin{equation}
    n_k = G_k n_{\mathrm{s}},
\end{equation}
where \( G_k \) is the channel gain, {\color{blue}which can be obtained through pilot-based or blind estimation methods~\cite{TCOM:2020Corr,TWC:2022Blind,PJ:2020ANN}.} We assume that \( G_k \) is known at both BS and users.  

From \eqref{eq:poissionPDF}, the Poisson distribution's defining property (e.g., the mean and variance are equal) introduces signal-dependent shot noise (SDSN) from the photon-conversion process, commonly referred to as Poisson shot noise. This noise exhibits nonlinear behavior resulting from the inherent nonlinearity of the Poisson distribution.

As the transmissions are uncoordinated and received asynchronously, we define a detection window with the duration of \( \tau L_{\mathrm{s}} \) for each user to align with and capture its frames at the receiver. This detection window can also capture frames from other users.
Suppose the beginnings of a user's frames are correctly estimated at the receiver. The receiver slides the detection window of the user with a step size of \( \tau L_{\mathrm{s}} \) to recover the signals of the user~\cite{TWC:jiangSC2022}. 

\subsection{Problem Statement}  
The goal is to design an effective and efficient receiver that can recover signals transmitted asynchronously (in a grant-free fashion) by an unknown number of users under Poisson shot noise.  
The key challenge is to identify the transmitting users and then estimate the starting points of frames transmitted by the users, as grant-free transmissions cause the frames to be received with different delays at the receiver. 

\begin{figure}
        \centering
        \includegraphics[width=0.88\columnwidth]{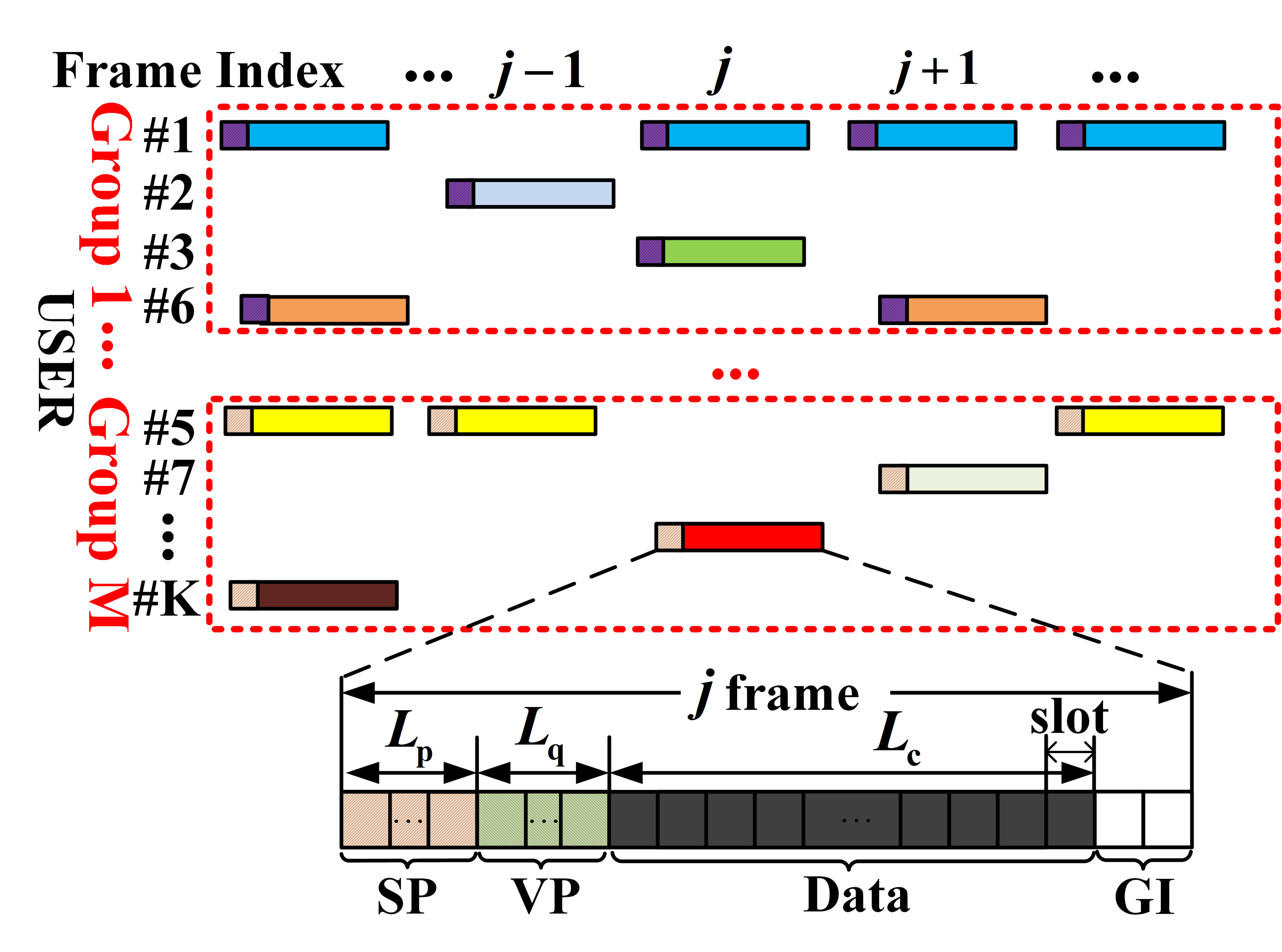}
        \caption{An illustration of the proposed signal frame structure for the grant-free uncoordinated
        transmission system.}
        \label{fig:FrameStructure}
    \end{figure}

\section{Delay Estimation for Grant-Free Transmissions}
\label{Sec:randomaccess}

\begin{figure}
        \centering
        \includegraphics[width=0.9\columnwidth]{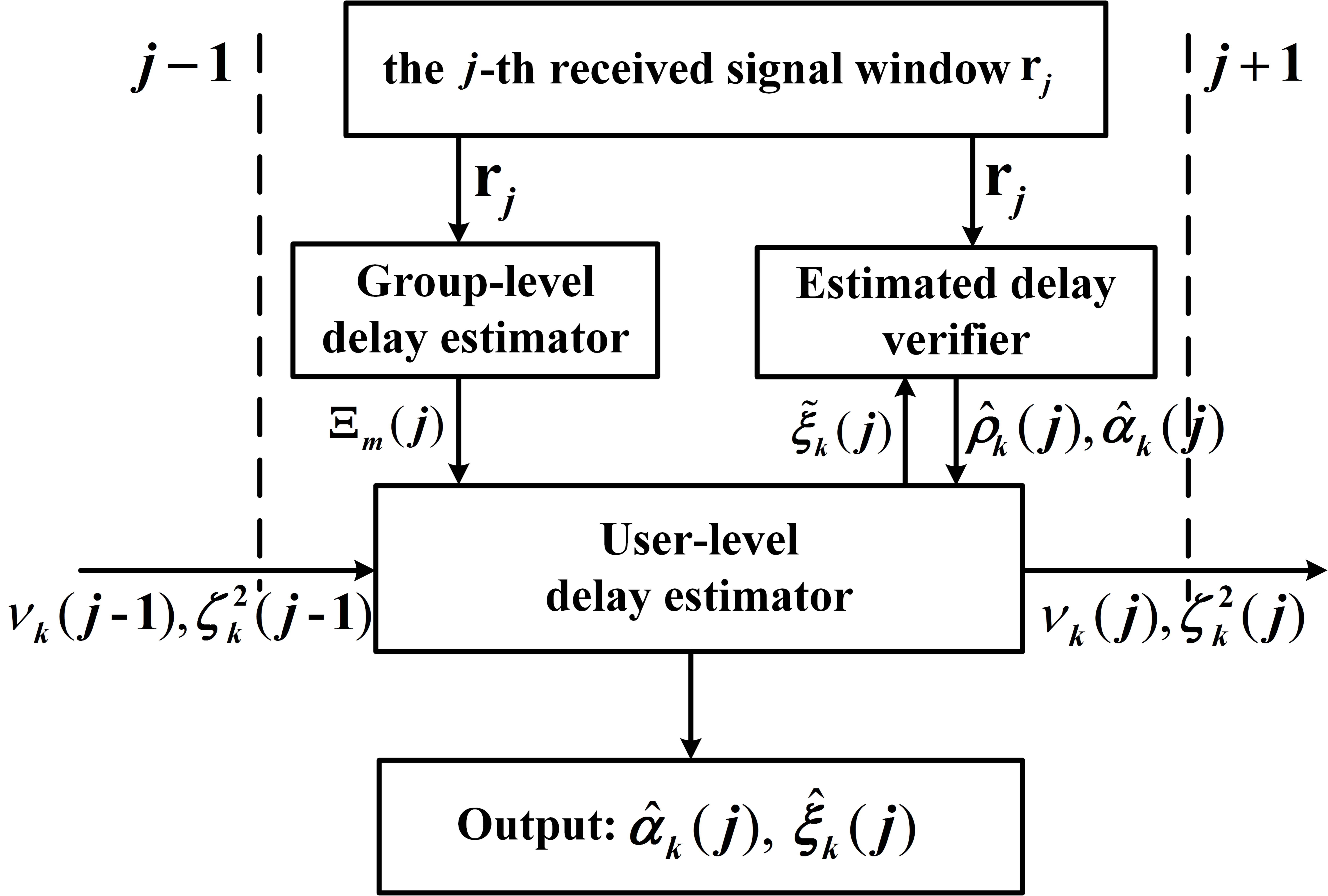}
        \caption{Flowchart of the proposed delay estimation and user activity estimation.}
        \label{fig:BayesianEstimation}
    \end{figure}

At the receiver, frame-level processing is employed to exploit Poisson statistical properties to estimate the effective delays of the received signals affected by Poisson shot noise, and synchronize the frames through three modules within a detection window, i.e., the $j$-th detection window for user $k$ who is in the $m$-th group:
\begin{itemize}
    \item \textit{Group-level delay estimator}: This module evaluates the correlation between the local SP and the received signals to determine the time delay for each group. It extracts delay values for all users in each group, collected by~$\Xi_{m}(j)$.
    
    \item \textit{User-level delay estimator}: For each user, this module estimates the time delay by maximizing the probability of the delay values within the set~$\Xi_{m}(j)$. The estimated delay for user $k$ is denoted as $\Tilde{\xi}_{k}(j)$.
    
    \item \textit{Delay verifier}: This module examines the estimated delays for each user based on the VP, providing verification factors $\hat{\rho}_{k}(j)$ to iteratively improve the accuracy of the delay estimates $\Tilde{\xi}_{k}(j)$. It also detects user activity using the VP. The verified time delay and activity for user $k$ in group $m$ are denoted as $\hat{\xi}_{k}(j)$ and $\hat{\alpha}_{k}(j)$, respectively.
\end{itemize}
Fig. \ref{fig:BayesianEstimation} presents the flowchart of this proposed synchronization method, with detailed operations for each module discussed in the following sections. 
        
    \subsection{Group-Level Delay Estimation}
    \label{sec:GL-estimator}
   
We estimate the time delay for each group by evaluating the correlation between the SP and the received signals in the $j$-th detection window, $\mathbf{r}_j = \{r_{j}(l), \, l = 1, \ldots, 3 L_{\mathrm{s}}\}$. Let $\mathbf{p}_m$ be the SP for the $m$-th group.

\subsubsection{Expected PhC Signals}  
    Using the received signal $\mathbf{r}_{j}$ in the $j$-th detection window, we estimate the expected signals using a denoising method. Background radiation is treated as noise. In the $l$-th slot of the detection window, the expected desired signals are given by
    \begin{equation}
         \Tilde{r}_{j} (l) = {r}_{j} (l) - n_{\mathrm{b}}.
    \end{equation}    
Then, we extract a symbol subsequence of length $L_{\mathrm{p}}$ starting from the $n$-th slot: ${\Tilde{r}_{j}(n+l), l = 1, \ldots, L_{\mathrm{p}}}$, which carries signal-dependent Poisson shot noise due to the PhC process.

\subsubsection{Detection of SP}  
    To detect the peak of the SP, we compute the autocorrelation of the pseudo-random sequence. For each $n \in [1, L_{\mathrm{s}}]$, we calculate the correlation between the symbol subsequence and the SP:
    \begin{equation}
    \label{eq:corre_A}
        R_{m,j}(n) = \frac{1}{L_{\mathrm{p}}} \sum_{l=1}^{L_{\mathrm{p}}} \Tilde{r}_{j}(n+l) \cdot p_{m}(l).
    \end{equation}       
    The value of $n$, at which $R_{m,j}(n)$ reaches its peak, corresponds to the location of the SP within the received signals. Since multiple users in group $m$ share the same SP, there can be multiple correlation peaks. The number of peaks is no more than that of users in group $m$. We consider all $n$ values with $R_{m,j}(n) > \varepsilon_{\mathrm{p}}$ as the group-level delay estimates for group $m$, where $\varepsilon_{\mathrm{p}}$ is the delay verification threshold. These values are stored in the set $\Xi_{m}(j)$, which is statistically influenced by signal-dependent Poisson shot noise in the received signals.

\vspace{2 mm}
\noindent \textbf{Remark:}
\textit{The group-level delay estimator reduces the number of users to be processed in the user-level delay estimators, 
thus improving accuracy and reducing complexity. 
This benefit arises from the reduction in computational burden; see Section~\ref{sec:complexity}. However, the number of user groups is constrained by the limited SPs available in engineering implementations. This trade-off balances efficiency and practical feasibility.}

    \subsection{User-Level Delay Estimation}
    \label{sec:UL-estimator}
 
Given the detection of group-level user delays, $\Xi_{m}(j)$ (see Sec.~\ref{sec:GL-estimator}), individual user delays within a group remain indistinguishable. To resolve this, each user's delay is identified by maximizing the PDF derived from observed delay values.

\subsubsection{PhC Delay Information}
The user-level delay estimator determines the delay for each user, i.e., $\Tilde{\xi}_{k}(j)$ in the $j$-th detection window for user $k$. Due to system clock offset, skewness, and drift, the random delays are known to follow a Gaussian distribution~\cite{kim:2011robust}. For user $k$, the delay in the $j$-th detection window follows
    ${\xi}_{k}(j) \sim \mathcal{N}\left(\mu_{k}(j), \sigma_{k}^{2}(j) \right)$,
where $\mu_{k}(j)$ and $\sigma_{k}^2(j)$ are the mean and variance, respectively. Let $u_{k}(j)$ and $\delta_{k}^{2}(j)$ denote the prior information's mean and variance, respectively. Updates of $u_{k}(j)$ and $\delta_{k}^2(j)$ are described in Section~\ref{App:UpdatesUserLevelDelayEstimator}.

\subsubsection{Delay Estimation}
Given prior information $u_{k}(j)$ and $\delta_{k}^{2}(j)$, the delay for user $k$ in group $m$ during the $j$-th window is estimated by maximizing the PDF of observed PhC delays. The PDF of ${\xi}_{k}(j)$ is 
    $p\left({\xi}_{k}(j)\right) = \frac{\exp\left[-\frac{\left({\xi}_{k}(j)-u_{k}(j)\right)^2}{2\delta_{k}^{2}(j)}\right]}{\sqrt{2\pi}{\delta_{k}(j)}}$.
Using the group-level delay estimator, possible delays for user $k$ are ${\xi}_{k}(j) \in {\Xi}_{m}(j)$. Let ${\Xi}_m\left[n_{\xi_k(j)} \right]$ be the $n_{\xi_k(j)}$-th element of ${\Xi}_{m}(j)$. When ${\xi}_{k}(j) = {\Xi}_m\left[n_{\xi_k(j)} \right]$, the PDF becomes
\begin{equation}
    \Pr\left({\xi}_{k}(j) = {\Xi}_m\left[n_{\xi_k(j)} \right]\right) = 
    \frac{\exp\left[-\frac{\left({{\Xi}_m\left[n_{\xi_k(j)} \right]} - u_{k}(j)\right)^2}{2\delta_{k}^{2}(j)}\right]}{\sqrt{2\pi}{\delta_{k}(j)}}.
\end{equation}
The estimated delay, $\Tilde{\xi}_{k}(j)$, is determined by maximizing 
\begin{equation}\label{eq:argxi}
    \Tilde{\xi}_{k}(j) = {\rho}_{k}(j) \times \!\!\!\!\!\! 
    \mathop{\arg \max}_{{{\Xi}_m\left[n_{\xi_k(j)} \right]} \in {\Xi}_{m}(j)} \!\!\!\!\!\!
    \Pr\left({\xi}_{k}(j) \!= {{\Xi}_m\!\!\left[n_{\xi_k(j)} \right]}\right),
\end{equation}
where $\Tilde{\xi}_{k}(j)$ is subject to signal-dependent Poisson shot noise, and ${\rho}_{k}(j)$ is the verification factor; see Section~\ref{sec:verifier}.

    \subsection{Estimated Delay Verification}
    \label{sec:verifier}
    \begin{figure*}[hbp]
    \hrulefill
        \begin{equation}
         \! \!\mathbf{r}_{k,j} \!=\! \big\{r_{k,j}({l}), l\!\in\!\left[1,L_{\mathrm{s}}\right]\big\}\! = \!\!
        \begin{cases}
        \!\big\{r_{k,{j-1}}( L_\mathrm{s}\!+\!\Tilde{\xi} _{k}(j\!-\!1) \!+\! 1),\ldots,r_{k,{j-1}}(L_\mathrm{s}), r_{k,j}(1),\ldots,r_{k,j}(L_\mathrm{s}\!+\!\Tilde{\xi}_{k}(j))\big\}, \!&\!\!\mathrm{if\ } \Tilde{\xi}_{k}(j) <0;\\        
        \!\big\{r_{k,j}(L_\mathrm{s}\!+\!\Tilde{\xi}_{k}(j)\!+\!1),\ldots,r_{k,j}(L_\mathrm{s}),r_{k,j+1}(1),\ldots,r_{k,j+1}(L_\mathrm{s}\!+\!\Tilde{\xi}_{k}(j\!+\!1))\big\}, &\!\!\mathrm{if\ } \Tilde{\xi}_{k}(j) \ge0.
        \end{cases}
        \label{eq:verifierReceivedSignal}
    \end{equation} 
    \end{figure*}

    
With the user-level delay $\Tilde{\xi}_{k}(j)$ estimated in Section~\ref{sec:UL-estimator}, the VP, $\mathbf{q}$, is utilized to validate the accuracy of these delays, addressing potential errors from PhC nonlinearity. 
As shown in Fig.~\ref{fig:SystemModel}, the VP is interleaved before transmission, requiring accurate delay estimation for proper deinterleaving. The delay estimation accuracy from Section~\ref{sec:UL-estimator} is verified by analyzing the VP in the received signals. 

\subsubsection{Extracting PhC VP Signals}

Using the estimated delay $\Tilde{\xi}_{k}(j)$ in \eqref{eq:argxi}, the corresponding received signal $\mathbf{r}_{k,j}$ for user $k$ in window $j$ is identified, as described in \eqref{eq:verifierReceivedSignal}. 
After deinterleaving the received signal $\mathbf{r}_{k,j}$ with the user-specific interleaver, 
the VP is extracted as $\mathbf{\Tilde{q}} = \{\Tilde{q}(l), l = 1, \ldots, L_{\mathrm{q}}\}$, which can be corrupted by MUI and noise.

\subsubsection{Verification of VP}

To validate the delay estimation, the correlation between the received VP, $\mathbf{\Tilde{q}} $, and the transmitted VP, $\mathbf{q}$, is computed using
\begin{equation}
    R_{k}(j) = \frac{1}{L_{\mathrm{q}}} \sum_{l=1}^{L_{\mathrm{q}}} \Tilde{q}(l) \cdot q(l). \label{eq:interVP}
\end{equation}
The verification factor $\hat{\rho}_{k}(j)$ is determined as
\begin{equation}
    \hat{\rho}_{k}(j) =  
    \begin{cases} 
        1, & \text{if } R_{k}(j) > \varepsilon_{\mathrm{q}}, \\ 
        0, & \text{if } R_{k}(j) \leq \varepsilon_{\mathrm{q}}, 
    \end{cases} \label{eq:calibrationFactor}
\end{equation}
where $\varepsilon_{\mathrm{q}}$ is the delay verification threshold.

\subsubsection{Activity Estimation} User activity estimation and delay verification jointly exploit the statistical properties of the Poisson process at the frame level. As shown in \eqref{eq:calibrationFactor}, if $\hat{\rho}_{k}(j) = 1$, the delay estimate $\Tilde{\xi}_{k}(j)$ from (11), which is affected by signal-dependent Poisson shot noise, is verified, and the final delay is output as $\hat{\xi}_{k}(j) = \Tilde{\xi}_{k}(j)$, indicating that the user is active, that is, $\hat{\alpha}_{k}(j) = 1$. If $\hat{\rho}_{k}(j) = 0$ after $T_{\mathrm{in}}$ verification rounds, it implies the absence of the VP, indicating $\hat{\alpha}_{k}(j) = 0$. The frame-level statistical approach can effectively estimate the delay through fewer rounds of verification, thereby mitigating the effect of Poisson shot noise and reducing computational complexity.


    \subsection{Synchronization Algorithm and Complexity Analysis}
    \label{sec:complexity}

    \begin{algorithm}[t]
\caption{Synchronization for grant-free PhC OWC}
\label{algo:BayesDelayEst}
\begin{algorithmic}[1]
    \State \textbf{Initialization:}
    \Statex \quad Set the maximum number of verification rounds $T_{\mathrm{in}}$;
    \Statex \quad Set the counter to 0;
    \State \textbf{Input:}
    \Statex \quad Parameters: $\mathbf{r}_{j}$, ${\mathbf{p}_m}$, ${\mathbf{q}}$, ${\mathbf{G}_m}$, $n_{\mathrm{b}}$, and $n_{\mathrm{s}}$, $\forall m$;
    \Statex \quad Prior information: $u_{k}(j)$, $\delta_{k}^2(j)$, $\forall k,j$;
    \State \quad Obtain ${\Xi}_{m}(j)$, $\forall m$, by comparing \eqref{eq:corre_A} and threshold~$\varepsilon_{\mathrm{p}}$;
    
    \While{$\hat{\rho}_{k}(j) = 0$ \textbf{and} $\mathrm{counter} < T_{\mathrm{in}}$}
        \State Obtain $\Tilde{\xi}_{k}(j)$ using \eqref{eq:argxi} and $\hat{\rho}_{k}(j)$ using \eqref{eq:calibrationFactor};
        \If{$\hat{\rho}_{k}(j) = 1$}
            \State Set $\hat{\xi}_{k}(j) = \Tilde{\xi}_{k}(j)$ and $\hat{\alpha}_{k}(j) = 1$;
        \EndIf
        \State $\mathrm{counter} \leftarrow \mathrm{counter} + 1$;
    \EndWhile
    
    \For{each $k$}
        \If{$\hat{\rho}_{k}(j) = 0$}
            \State Set $\hat{\alpha}_{k}(j) = 0$;
        \EndIf
    \EndFor
    
    \State \textbf{Output:}
    \Statex \quad Estimation results: $\hat{\alpha}_{k}(j)$, $\hat{\xi}_{k}(j)$, $\forall k,j$;
    \Statex \quad Posterior information: $\nu_{k}(j)$, $\zeta_{k}^2(j)$, $\forall k,j$.
\end{algorithmic}
\end{algorithm}
    
Based on the group-level delay estimator, user-level delay estimator, and delay verifier discussed in Sections \ref{sec:GL-estimator} to \ref{sec:verifier}, the proposed synchronization algorithm is summarized in \textbf{Algorithm~\ref{algo:BayesDelayEst}}. Step 3 performs group-level delay estimation, followed by user-level delay estimation in Step 5. Steps 6–16 verify the delay estimation results and detect user activity.

The computational complexity of \textbf{Algorithm \ref{algo:BayesDelayEst}} is determined by the group-level delay estimator, user-level delay estimator, and delay verifier.

\begin{enumerate}
    \item \textit{Group-Level Delay Estimator:} 
    The computational cost is primarily influenced by the correlation process, which scales as $\mathcal{O}(L_{\mathrm{p}}L_{\mathrm{s}})$ per user. For $K$ users, the overall complexity becomes $\mathcal{O}(KL_{\mathrm{p}}L_{\mathrm{s}})$.

    \item \textit{User-Level Delay Estimator:} 
    Let $K_{\mathrm{a}}$ denote the number of potential delay values (or active users) in group $m$. The user-level estimation computes probabilities for each user, scaling as $\mathcal{O}(K_{\mathrm{a},m}K_{m})$, where $K_{m}$ is the total number of users in group $m$. 
    User delay estimation is achieved by maximizing the PDF of the observed values. Since this process can be parallelized across groups, the time complexity reduces to $\mathcal{O}[\max_{\forall m}(K_{\mathrm{a},m}K_{m})]$.

    \item \textit{Estimated Delay Verifier:} 
    The complexity of verifying a user's delay scales as $\mathcal{O}(L_{\mathrm{q}})$. For all users in a group, parallel verification yields the total complexity of $\mathcal{O}[\max_{\forall m}(MK_{m}L_{\mathrm{q}})]$.
\end{enumerate}
Consider the iterative nature of the algorithm, which performs $T_{\mathrm{in}}$ rounds of user-level delay estimation and delay verification (see Fig. \ref{fig:FrameStructure}). The overall computational complexity is $\mathcal{O}\{KL_{\mathrm{p}}L_{\mathrm{s}} + T_{\mathrm{in}}[\max_{\forall m}(K_{\mathrm{a},m}K_{m}) + \max_{\forall m}(MK_{m}L_{\mathrm{q}})]\}$.

The group-level delay estimation significantly reduces computational complexity by enabling parallel user-level estimation and verification processes across groups. Without this grouped approach, the algorithm would need to estimate and verify delays for each user individually, leading to a higher computational burden of up to $\mathcal{O}[T_{\mathrm{in}}(KL_{\mathrm{p}}L_{\mathrm{s}} + K_{\mathrm{a}}K + KL_{\mathrm{q}})]$, where $K_{\mathrm{a}}$ represents the total number of active users.
{\color{blue}Table~\ref{TableComplexity} provides a breakdown of computational complexity. To further demonstrate the efficiency gain of the grouped design, we further provide a runtime comparison based on MATLAB simulation. The experiments were performed on a ThinkPad T14 laptop (Intel i7-10510U CPU, 16 GB RAM) using MATLAB R2018b. For one frame of processing, the grouped design and ungrouped design require 0.131~s and 0.523~s, respectively, indicating an approximately 75\% reduction in runtime.}

    \begin{table} 
    \centering
    \caption{Computational Complexity of User Delay Estimation and User Activity Using Synchronization Algorithm.}
    \label{TableComplexity}
		\begin{tabular}  
        {lcc}  
			\toprule
			\quad 
			& Multiple SP (grouped) 
			& Single SP (ungrouped)\\
			\midrule
			Group-level delay esti.  
			& $ K L_{\mathrm{p}}  L_{\mathrm{s}}$ 
			& --\\
			
			User-level delay esti.
			& $\max\limits_{\forall m}\left(K_{m} K_{\mathrm{a},m}\right)$ 
			& $ K L_{\mathrm{p}}  L_{\mathrm{s}} + K K_{\mathrm{a}}$\\

            Estimated delay veri.
            & $ \max\limits_{\forall m}\left(M K_{\mathrm{m}} L_\mathrm{q}\right) $ 
			& $K L_\mathrm{q}$\\
			\bottomrule
		\end{tabular}				 
	\end{table}

    \subsection{CRLB for Delay Estimation under PhC Noise}
    To analyze the achievable accuracy of delay estimation under a signal-dependent Poisson model, we derive the Cramer–Rao lower bound (CRLB) for unbiased estimators. The CRLB provides the minimum variance achievable by any unbiased estimator of the delay parameter, thereby serving as a benchmark for estimator performance~\cite{BOOK:van2004detection}. 
    It is determined by the Fisher information obtained from the log-likelihood function of the observed data. The Fisher information quantifies the expected sensitivity of the likelihood function with respect to the parameter of interest, and thus reflects the information contained in the data about the unknown delay~\cite{BOOK:van2004detection}.

    \subsubsection{PhC Observation Model}
    Let ${{\mathbf{r}}_{k,j}}=\{r_{k,j}(l), \, l = 1, \ldots, L_{\mathrm{s}}\}$ denote the photon counts observed for user $k$ in the $j$-th detection window. 
    The likelihood function of the unknown delay parameter ${{\xi }_{k}}(j)$ is  $\Pr \left( {{\mathbf{r}}_{k,j}}\mid {{\xi }_{k}}(j) \right)$, with each observation following the Poisson distribution in~\eqref{eq:poissionPDF}. To derive CRLB, we assume that the transmitted symbol sequence is known at the receiver~\cite{TCOM:2004Dabora}. In this case, the delayed received waveform for user $k$ in window $j$ is written as $s_{k,j}(l-\xi_k(j))$ and modeled as a root-raised-cosine pulse~\cite{AO:2008Bound}. With long, well-correlated pilot sequences, multi-user interference becomes predictable and can be treated as known to derive the single-user CRLB.

    The log-likelihood function is given by
    \begin{subequations}
    \begin{align} 
    \ln \Pr ({\mathbf{r}_{k,j}};{\xi _k}(j)) 
    &= \sum\limits_{l = 1}^{{L_{\mathrm{s}}}} {\ln \Pr ({r_{k,j}}(l);{\xi _k}(j))} \label{eq:poisson_observation_1}\\
    &= \sum\limits_{l = 1}^{{L_{\mathrm{s}}}} \Big[ {r_{k,j}}(l)\ln n_{k,j}(l;{\xi _k}(j)) \nonumber\\
    & \ \ \ \ - n_{k,j}(l;{\xi _k}(j)) - \ln \big( {r_{k,j}}(l)! \big) \Big] , \label{eq:poisson_observation_model} 
    \end{align}
    \end{subequations} 
    where \eqref{eq:poisson_observation_model} is obtained by substituting (3) into \eqref{eq:poisson_observation_1}.
    
    Differentiating the log-likelihood with respect to $\xi_k(j)$ yields the score function
    \begin{subequations}
    \begin{flalign}
    {\mathbb{S}}&({{\mathbf{r}}_{k,j}};{\xi _k}(j)) 
     = \frac{{\partial \ln \Pr ({{\mathbf{r}}_{k,j}};{\xi _k}(j))}}{{\partial {\xi _k}(j)}} \label{eq:first_order_1} \\
    & = \sum\limits_{l = 1}^{{L_{\mathrm{s}}}} {\left( {\frac{{{r_{k,j}}(l)}}{{{n_{k,j}}(l;{\xi _k}(j))}} - 1} \right) \cdot \frac{{\partial {n_{k,j}}(l;{\xi _k}(j))}}{{\partial {\xi _k}(j)}}}, \label{eq:first_order_3} 
    \end{flalign}
    \end{subequations}
    where \eqref{eq:first_order_3} is obtained by substituting (13b) into \eqref{eq:first_order_1} and simplifying it by taking the derivative.

    \subsubsection{Fisher Information}
    Given $\mathbb{S}({{\mathbf{r}}_{k,j}};{\xi _k}(j))$, the Fisher information is the second moment of the score, as given by
    \begin{subequations}
    \begin{flalign}
    & {\cal I}({\xi _k}(j)) = \mathbb{E}\big[ {{{\left( {\mathbb{S}({{{\mathbf{r}}}_{k,j}};{\xi _k}(j))} \right)}^2}} \big] \label{eq:Fisher_1}\\
    & = \mathbb{E}\Bigg[ {{{\Big( {\sum\limits_{l = 1}^{{L_{\mathrm{s}}}} {\big( {\frac{{{r_{k,j}}(l)}}{{n_{k,j}(l;{\xi _k}(j))}} - 1} \big)}  \cdot \frac{{\partial {n_{k,j}}(l;{\xi _k}(j))}}{{\partial {\xi _k}(j)}}} \Big)}^2}} \Bigg] \label{eq:Fisher_2}\\
    & = \mathbb{E}\Bigg[ \sum\limits_{l = 1}^{{L_{\mathrm{s}}}} {\sum\limits_{m = 1}^{{L_{\mathrm{s}}}} {\Big( {\frac{{{r_{k,j}}(l)}}{{{n_{k,j}}(l;{\xi _k}(j))}} - 1} \Big)\Big( {\frac{{{r_{k,j}}(m)}}{{{n_{k,j}}(m;{\xi _k}(j))}} - 1} \Big)} } \nonumber \\
    & \ \ \ \ \ \ \ \ \ \ \ \ \ \ \ \ \frac{{\partial {n_{k,j}}(l;{\xi _k}(j))}}{{\partial {\xi _k}(j)}}\frac{{\partial {n_{k,j}}(m;{\xi _k}(j))}}{{\partial {\xi _k}(j)}} \Bigg] , \label{eq:Fisher_3}
    \end{flalign}
    \end{subequations}
    where \eqref{eq:Fisher_2} is obtained by substituting \eqref{eq:first_order_3} into \eqref{eq:Fisher_1}, and the simplified to \eqref{eq:Fisher_3}.

    Expanding and separating diagonal and cross terms yields
    \begin{align}
    & {\cal I}({\xi _k}(j)) = 
    \sum\limits_{l = 1}^{{L_{\mathrm{s}}}} \mathbb{E}{\Big[ {{{\Big( {\frac{{{r_{k,j}}(l)}}{{{n_{k,j}}(l;{\xi _k}(j))}} - 1} \Big)}^2}} \Big]} {\Big( {\frac{{\partial {n_{k,j}}(l;{\xi _k}(j))}}{{\partial {\xi _k}(j)}}} \Big)^2} \nonumber\\
    & \ \ \ \ \ \ \ \ \ \ + \sum\limits_{l \ne m}^{{L_{\mathrm{s}}}} \mathbb{E}{\Big[ {\Big( {\frac{{{r_{k,j}}(l)}}{{{n_{k,j}}(l;{\xi _k}(j))}} - 1} \Big)\Big( {\frac{{{r_{k,j}}(m)}}{{{n_{k,j}}(m;{\xi _k}(j))}} - 1} \Big)} \Big]} \nonumber\\
    & \ \ \ \ \ \ \ \ \ \ \ \ \ \ \ \ \frac{{\partial {n_{k,j}}(l;{\xi _k}(j))}}{{\partial {\xi _k}(j)}}\frac{{\partial {n_{k,j}}(m;{\xi _k}(j))}}{{\partial {\xi _k}(j)}}. \label{eq:Fisher_4}
    \end{align}
    Since Poisson observations are independent across time slots, the cross terms vanish. Using $\mathbb{E} \left[ {{r_{k,j}}(l)} \right] = {n_{k,j}}(l;{\xi _k}(j))$ and $\mathbb{E}\big[ {{{\big( {{r_{k,j}}(l)} \big)}^2}} \big] = {n_{k,j}}(l;{\xi _k}(j)) + {\big( {{n_{k,j}}(l;{\xi _k}(j))} \big)^2}$~\cite{BOOK:2007probability}, the expectation simplifies to
    \begin{subequations}
    \begin{align}
    &\!\!\!\!\mathbb{E}\Big[ {{\big( {\frac{{{r_{k,j}}(l)}}{{{n_{k,j}}(l;{\xi _k}(j))}} \!-\! 1} \big)^2}} \Big] \nonumber  \\
    &\!\!\!\!= \mathbb{E}\Big[ \frac{{{{\big( {{r_{k,j}}(l)} \big)}^2}}}{{{{\big( {{n_{k,j}}(l;{\xi _k}(j))} \big)}^2}}} - \frac{{2{r_{k,j}}(l)}}{{{n_{k,j}}(l;{\xi _k}(j))}} + 1  \Big] \label{eq:Fisher_temp2_1}\\
    &\!\!\!\! =\! \frac{{{n_{k,j}}(l;{\xi _k}(j))\!\! + \!\!{{\big( {{n_{k,j}}(l;{\xi _k}(j))} \big)}^2}}}{{{{\big( {{n_{k,j}}(l;{\xi _k}(j))} \big)}^2}}} \! -\! \frac{{2{n_{k,j}}(l;{\xi _k}(j))}}{{{n_{k,j}}(l;{\xi _k}(j))}} \!\!+ \!\!1 \label{eq:Fisher_temp2_2}\\
    &\!\!\!\! = \frac{1}{{{n_{k,j}}(l;{\xi _k}(j))}},\label{eq:Fisher_temp2_3}
    \end{align}
    \end{subequations}
    where \eqref{eq:Fisher_temp2_1} is obtained by square expansion, \eqref{eq:Fisher_temp2_2} is obtained by simplifying \eqref{eq:Fisher_temp2_1} using the Poisson property, and \eqref{eq:Fisher_temp2_3} is obtained by simplification through addition and subtraction.
    
    By plugging \eqref{eq:Fisher_temp2_3} in \eqref{eq:Fisher_4}, the Fisher information becomes
    \begin{equation}
        {\cal I}({\xi _k}(j)) 
     = \sum\limits_{l = 1}^{{L_{\mathrm{s}}}} {\frac{1}{{{n_{k,j}}(l;{\xi _k}(j))}}} {\Big( {\frac{{\partial {n_{k,j}}(l;{\xi _k}(j))}}{{\partial {\xi _k}(j)}}} \Big)^2} \label{eq:Fisher_temp3_1}.
    \end{equation}
    Since ${n_{k,j}}(l;{\xi _k}(j)) = {n_{\mathrm{b}}} + {{n_{\mathrm{s}}}{s_{k,j}}(l - {\xi _k}(j))} $, we obtain
    \begin{equation}
        \frac{{\partial {n_{k,j}}(l;{\xi _k}(j))}}{{\partial {\xi _k}(j)}} =  - {{n_{\mathrm{s}}}{s_{k,j}^{\prime}}(l - {\xi _k}(j))}. \label{eq:Fisher_temp4_1}
    \end{equation}
    Plugging \eqref{eq:Fisher_temp4_1} into \eqref{eq:Fisher_temp3_1}, the Fisher information is updated to
    \begin{equation}
        {\cal I}({\xi _k}(j)) = \sum\limits_{l = 1}^{{L_{\mathrm{s}}}} {\frac{{{{\big( { {{n_{\mathrm{s}}} s_{k,j}^{\prime} (l - {\xi _k}(j))} } \big)}^2}}}{{{n_{\mathrm{b}}} + {{n_{\mathrm{s}}}{s_{k,j}}(l - {\xi _k}(j))} }}}. \label{eq:Fisher_temp5_1}
    \end{equation}
    
    \subsubsection{PhC CRLB}
    Under standard regularity conditions, the variance of any unbiased estimator $\hat\xi_k(j)$ is lower bounded by the inverse of the Fisher information. Specifically, under the photon-counting model, the CRLB takes the following form: 
    \begin{subequations}
        \begin{align}
        {\mathrm{Var}}({\hat \xi _k}(j)) & \ge \frac{1}{{{\cal I}({\xi _k}(j)) }} \\
        & = {\Bigg[ {\sum\limits_{l = 1}^{{L_{\mathrm{s}}}} {\frac{{{{\big( { {{n_{\mathrm{s}}} s_{k,j}^{\prime} (l - {\xi _k}(j))} } \big)}^2}}}{{{n_{\mathrm{b}}} + {{n_{\mathrm{s}}}{s_{k,j}}(l - {\xi _k}(j)) } }}} } \Bigg]^{ - 1}}.
        \end{align}
    \end{subequations}
    This bound explicitly shows the dependence of the achievable delay estimation accuracy on the $n_{\mathrm{s}}$ and $n_{\mathrm{b}}$.

    \section{Iterative MUD with Delay Calibration}
    \label{sec:Iterative detection}
The nonlinear characteristics of Poisson shot noise result in the mixing of signals and noise in the received signals, creating challenges for accurate user delay estimation. To address this, a Bayesian method is employed to enhance the accuracy of prior delay updates, improving the delay estimation described in Section \ref{Sec:randomaccess}. The estimated delays serve as inputs to the proposed MUD for asynchronous, grant-free transmissions.

Unlike the traditional iterative MUD designed for synchronized transmissions~\cite{OE:zhouXL2012}, our algorithm leverages delay information to develop an iterative MUD scheme for asynchronous, grant-free transmissions. 
Bayesian delay updates are used to refine prior delay estimates, improving the accuracy of delay estimation. A detection window is established based on the delays estimated using \textbf{Algorithm~\ref{algo:BayesDelayEst}}, mitigating the misalignment between asynchronously received frames and accurately identifying interfering symbols. 
The system employs slot-by-slot iterative detection using the principle of maximum \textit{a-posteriori} probability to restore the original symbols.

\subsection{Bayesian Delay Estimation}\label{App:UpdatesUserLevelDelayEstimator}    

To estimate the delay affected by Poisson noise using a Bayesian approach, we establish in Section \ref{sec:UL-estimator} that ${\xi}_{k}$ follows a Gaussian distribution: ${\xi}_{k} \sim \mathcal{N}({\xi}_{k}(j), \varphi_{k}^2)$. 
Then, we estimate the delay parameter ${\xi}_{k}(j)$. This involves determining its posterior distribution, conditioned on~${\xi}_{k}$. To do so, we define the prior distribution, derive the likelihood from prior and received signals, and the posterior distribution.

\subsubsection{Prior Distribution}
The prior distribution of the delay ${\xi}_{k}(j)$ is Gaussian: ${\xi}_{k}(j)  \sim \mathcal{N}(u_{k}(j), \delta_{k}^{2}(j))$,  where $u_{k}(j)$ and $\delta_{k}^{2}(j)$ represent the mean and variance of the previous detection window, and they are known parameters. 
The prior distribution of the delay can be updated on the basis of detection windows. The mean $\nu_{k}(j)$ and variance $\zeta_{k}^2(j)$ of the posterior distribution from the previous detection window are utilized as the mean and variance of the prior distribution for the current window, respectively. This update is given by 
    \begin{subequations}   
    \begin{align}
        u_{k}(j) = \nu_{k}(j-1), \\ 
        \delta_{k}^{2}(j)=\zeta_{k}^2(j-1),
        \label{eq:Updatemean1}
    \end{align}
    \end{subequations}
    where $u_{k}(1)$ is initialized to the center of the expected maximum delay search range, and $\delta_{k}^{2}(1)$ is set to 1, considering system clock accuracy and receiver hardware jitter.

\subsubsection{Likelihood Function} Using the estimated user delay $\tilde{\xi}_{k}(j)$ from \eqref{eq:argxi} and the posterior mean from previous detection windows, we estimate the likelihood function by averaging over the delays. Let $p({\xi}_{k} | {\xi}_{k}(j))$ denote the likelihood function, representing the PDF of ${\xi}_{k}$ conditioned on ${\xi}_{k}(j)$.

The mean $\upsilon_{k}(j)$ and variance $\phi_{k}^2(j)$ of the likelihood function, following a Gaussian distribution, are updated as 
    \begin{subequations}
    \begin{align}
    \label{eq:updateMean}
        {\upsilon_{k}(j)} 
        &= \mathbb{E}\left[ { \left({\hat{\xi}_{k}(j'), \forall j' < j }\right),{\tilde {\xi}_{k}(j)}} \right] \nonumber\\
        &= \mathbb{E}\left[ { \left({\nu_{k}(j'), \forall j' < j }\right),{\tilde {\xi}_{k}(j)}} \right];\\
    \label{eq:updateVar}
        \phi_{k}^2(j) &= {\mathop{\mathrm {Var}}} \left[ { \left({\nu_{k}(j')}, \forall j' < j \right),{\tilde {\xi}_{k}(j)}} \right],
    \end{align}
    \end{subequations}
where $\left({\nu_{k}(j'), \forall j' < j }\right)$ are the set of posterior means of the previous ${\left(j-1\right)}$ detection windows. $\mathbb{E}[\cdot]$ and $\mathop{\mathrm{Var}}[\cdot]$ represent mean and variance, respectively. $\tilde{\xi}_{k}(j)$ is affected by Poisson noise, and so is the estimated likelihood function.

\subsubsection{Joint Distribution} The joint distribution is derived as
    \begin{equation}
        p\left( {\xi}_{k}, {\xi}_{k}(j) \right) \!=\! p \left( {\xi}_{k}(j) \right) p \left( {\xi}_{k} \left| {\xi}_{k}(j) \right. \right),   \label{eq:jointDistribution}
    \end{equation}
    which follows from the joint distribution definition. The joint distribution can also be expressed as
    \begin{subequations}
        \begin{flalign}
            p\left( {\xi}_{k}, {\xi}_{k}(j) \right) 
            &\!=\!p\left( {\xi}_{k}(j), {\xi}_{k} \right) \label{eq:jointDistributionexchangeA}\\
            &\!=\!p \left( {\xi}_{k}(j)\left| {\xi}_{k} \right. \right) p\left( {\xi}_{k} \right) , \label{eq:jointDistributionexchangeB}
        \end{flalign}
    \end{subequations}
where \eqref{eq:jointDistributionexchangeA} uses the multiplication rule of probabilities, and \eqref{eq:jointDistributionexchangeB} defines the joint distribution. The marginal PDF, $p({\xi}_{k})$, is obtained by integrating over ${\xi}_{k}(j)$:
\begin{equation}
        p\left( {\xi}_{k} \right) =\smallint _{\varTheta }p \left( {\xi}_{k},\xi _{k}(j) \right) d\xi _{k}(j).
        \label{eq:marginalPDF}
    \end{equation}    
By integrating the joint PDF, we obtain the PDF, $p({\xi}_{k})$, of the random variable ${\xi}_{k}$.

\subsubsection{Posterior Distribution}Given the joint distribution given in~\eqref{eq:jointDistribution} to \eqref{eq:marginalPDF}, the posterior distribution $p(\xi_{k}(j) | {\xi}_{k})$ is 
\begin{subequations}
\begin{flalign}
p(\xi_{k}(j) | {\xi}_{k}) &= \frac{p({\xi}_{k}, \xi_{k}(j))}{p({\xi}_{k})} \label{eq:posteriorinformationA} \\
&= \frac{p(\xi_{k}(j)) p({\xi}_{k} | \xi_{k}(j))}{\int_{\varTheta} p({\xi}_{k}, \xi_{k}(j)) d\xi_{k}(j)} \label{eq:posteriorinformationB} \\
&= c \cdot p(\xi_{k}(j)) p({\xi}_{k} | \xi_{k}(j)) \label{eq:posteriorinformationC} \\
&\propto p(\xi_{k}(j)) p({\xi}_{k} | \xi_{k}(j)), \label{eq:postPDF}
\end{flalign}
\end{subequations}
where \eqref{eq:posteriorinformationA} is derived from \eqref{eq:jointDistributionexchangeB}, \eqref{eq:posteriorinformationB} comes from substituting \eqref{eq:jointDistribution} and \eqref{eq:marginalPDF} into \eqref{eq:posteriorinformationA}, and in \eqref{eq:posteriorinformationC}, $p({\xi}_{k})$ is treated as a constant $c$ since it only integrates over $\xi_{k}(j)$. Hence, \eqref{eq:postPDF} indicates that the posterior distribution is proportional to the prior distribution multiplied by the likelihood function.

Since both $p(\xi_{k}(j))$ and $p({\xi}_{k} | \xi_{k}(j))$ are Gaussian PDFs, they can be expressed as
\begin{equation}
p(\xi_{k}(j)) \propto \exp \left[ -\frac{(\xi_{k}(j) - u_{k}(j))^2}{2 \delta_{k}^{2}(j)} \right], \label{eq:priorPDFb}
\end{equation}
\begin{equation}
p({\xi}_{k} | \xi_{k}(j)) \propto \exp \left[ -\frac{({\xi}_{k} - \upsilon_{k}(j))^2}{2 \phi_{k}^2(j)} \right]. \label{eq:likePDF}
\end{equation}

Substituting \eqref{eq:priorPDFb} and \eqref{eq:likePDF} into \eqref{eq:postPDF} yields:
\begin{subequations}
\begin{flalign}
p(\xi_{k}(j) | {\xi}_{k}) \!
&\propto\! \exp \!\left[ -\frac{(\xi_{k}(j) \!\!- \!\!u_{k}(j))^2}{2 \delta_{k}^{2}(j)} - \frac{({\xi}_{k}\!\! - \!\!\upsilon_{k}(j))^2}{2 \phi_{k}^2(j)}\! \right] \\
&= \!\exp\! \left[ -\frac{(\xi_{k}(j) \!\!- \!\!u_{k}(j))^2}{2 \delta_{k}^{2}(j)}\! - \!\frac{({\xi}_{k} \!\!- \!\!\upsilon_{k}(j))^2}{2 \phi_{k}^2(j)} \!\right]. \label{eq:posteriorlast}
\end{flalign}
\end{subequations}

Since both $\xi_{k}(j)$ and $\upsilon_{k}(j)$ are influenced by nonlinear Poisson shot noise and MUI, the numerator and denominator of \eqref{eq:posteriorlast} are affected. Let $\nu_{k}(j)$ and $\zeta_{k}^2(j)$ denote the mean and variance of \eqref{eq:posteriorlast}. The updated values for $\nu_{k}(j)$ and $\zeta_{k}^2(j)$ are:
\begin{subequations}   
\begin{align}
\nu_{k}(j)\! &= \!\frac{u_{k}(j) \phi_{k}^2(j)\!\! +\!\! \upsilon_{k}(j) \delta_{k}^{2}(j)}{\delta_{k}^{2}(j) \!\!+\!\! \phi_{k}^2(j)}, \label{eq:updatePostMean} \\
\zeta_{k}^2(j) &= \frac{\delta_{k}^{2}(j) \phi_{k}^2(j)}{\delta_{k}^{2}(j) + \phi_{k}^2(j)}. \label{eq:updatePostVar}
\end{align}
\end{subequations}


\subsection{Detection Window} \label{sec:IUC-estimator}
\begin{figure}
        \centering
        \includegraphics[width=0.92\columnwidth]{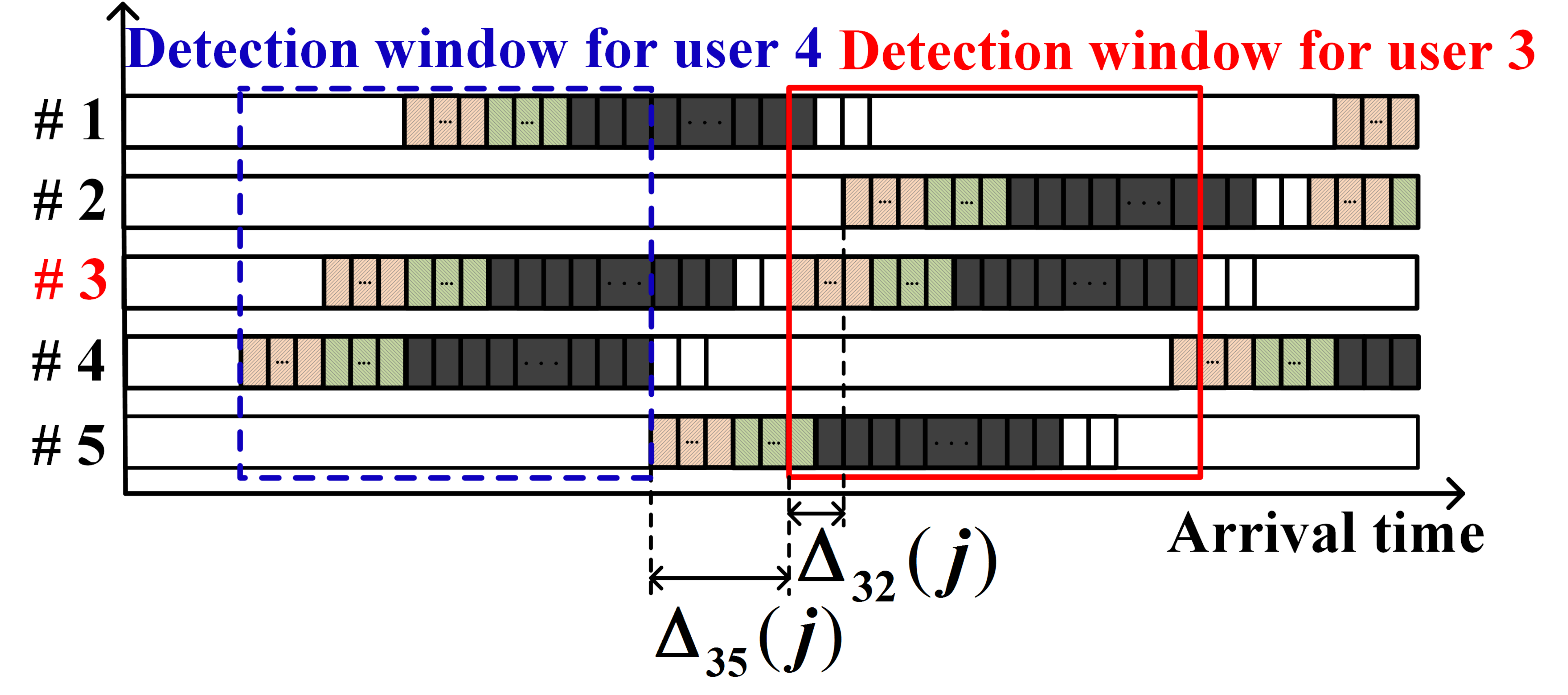}
        \caption{An example of the detection window for user 3 at the receiver of the considered, grant-free, multiuser, underwater PhC OWC system.}
        \label{fig:FrameWindow}
\end{figure}

With the prior delay $u_{k}(j)$ updated as in Section \ref{App:UpdatesUserLevelDelayEstimator} and the delay $\hat{\xi}_{k}(j)$ estimated in Section \ref{Sec:randomaccess}, symbol-level fine-tuning within the detection window enables more precise estimation of interfering symbols.
As shown in Fig. \ref{fig:FrameWindow}, with the length $L_{\mathrm{s}}$, of a detection window, 
the received frame within the window synchronized with the $j$-th frame of user~$k$ is 
\begin{equation}
    \mathbf{r}_{k, j} \!= \!\left[ r_{k, j}(l), l \!=\! 1, \!\ldots,\! L_{\mathrm{s}} \right] \!=\! \left[ r_{l(k, j)}, \!\ldots,\! r_{l(k, j) \!+ \!L_{\mathrm{s}} \!- \!1} \right],
\end{equation} 
where \( l(k, j) \) is the index of the slot marking the start of the $j$-th frame of user $k$.

Let \({\Tilde{s}}_{kk',j}(l)\) be the interfering symbol from user \( k' \) affecting the \( l \)-th slot/symbol in the \( j \)-th frame of user \( k \). Owing to the asynchronous, grant-free transmissions of the users, interference in the \( j \)-th frame of user \( k \) may arise from transmissions in the \((j-1)\)-th, \( j \)-th, or \((j+1)\)-th frame of other users. Therefore, \({\Tilde{s}}_{kk',j}(l)\) is given by
\begin{subnumcases}
        {{\Tilde{s}}_{kk',j}(l) \!\! = \!\!}   
        \! s_{k',j-1}(l - \Delta _{kk'}(j) + {L_{\mathrm{s}}})\!,\!  & \nonumber\\        &\!\!\!\!\!\!\!\!\!\!\!\!\!\!\!\!\!\!\!\!\!\!\!\!\!\!\!\!\!\!\!\!\!\!\!\!\!\!\!\!\!\!\!\!\!\!\!\!\!\!\!\!\!\!\!\!\!\!\!\!\!\!\!\!\!\!\!\!{\rm{for\ }}$\Delta _{kk'}(j) \!\ge\! 0 {\rm{\ and\ }}\!\! -\! \!{L_{\mathrm{s}}} \!<\! l \!-\! \Delta _{kk'}(j) \!\le\! 0$;\label{eq:interA} \\
        \! s_{k',j}(l - \Delta _{kk'}(j)),\! & \nonumber\\
        & \!\!\!\!\!\!\!\!\!\!\!\!\!\!\!\!\!\!\!\!\!\!\!\!\!\!\!\!\!\!\!\!\!\!\!\!\!\!\!\!\!\!\!\!\!\!\!\!\!\!\!\!\!\!\!\!\!\!\!\!\!\!\!\!\!\!\!\!{\rm{for\ }}$\Delta _{kk'}(j) \!\ge\! 0${\rm{\ and\ }} $l- \Delta _{kk'}(j) > 0$; \label{eq:inter_oldB} \nonumber\\        &\!\!\!\!\!\!\!\!\!\!\!\!\!\!\!\!\!\!\!\!\!\!\!\!\!\!\!\!\!\!\!\!\!\!\!\!\!\!\!\!\!\!\!\!\!\!\!\!\!\!\!\!\!\!\!\!\!\!\!\!\!\!\!\!\!\!\!\!{\rm{or\ }} $\Delta _{kk'}(j) \!<\! 0${\rm{\ and\ }} $l - \Delta _{kk'}(j) \le {L_{\rm{s}}}$; \label{eq:inter_oldA}\\
        \! s_{k',j+1}(l - \Delta _{kk'}(j) - {L_{\rm{s}}})\!,\! &  \nonumber\\        &\!\!\!\!\!\!\!\!\!\!\!\!\!\!\!\!\!\!\!\!\!\!\!\!\!\!\!\!\!\!\!\!\!\!\!\!\!\!\!\!\!\!\!\!\!\!\!\!\!\!\!\!\!\!\!\!\!\!\!\!\!\!\!\!\!\!\!\!\!{\rm{for\ }}$\Delta _{kk'}(j) \!\!<\! 0${\rm{\ and\ }} \!\!${L_{\rm{s}}} \!<\! l \!-\! \Delta _{kk'}(j) \!\!<\! 2{L_{\rm{s}}}$,
        \label{eq:interB}
    \end{subnumcases}
where $s_{k',j-1}(.)$ is the signal of user $k'$ in the $(j-1)$-th frame, \(\Delta_{kk'}(j) = \xi_{k'}(j) - \xi_{k}(j)\) denotes the relative delay between users \( k \) and \( k' \), and \(\xi_{k}(j)\) is the effective delay between user \( k \) and the receiver. Clearly, the users are synchronized in the system if \(\xi_{k}(j) = 0\). Here, 
\eqref{eq:interA} accounts for interference caused by the transmission of the \((j-1)\)-th frame of user \( k' \).  
\eqref{eq:inter_oldB} captures interference from the transmission of the \( j \)-th frame of user \( k' \).  
\eqref{eq:interB} indicates interference from the transmission of the \((j+1)\)-th frame of user \( k' \).  

\subsection{Iterative MUD for Asynchronous Transmissions}
\label{sec:AMUI-Detector}
    
Given the interfering symbols identified in Section~\ref{sec:IUC-estimator}, we proceed to propose the iterative MUD to detect signals. This method iteratively refines the identification of interfering symbols estimated in Section \ref{sec:IUC-estimator} for enhanced accuracy. Specifically, for the grant-free, underwater PhC OWC system, the proposed iterative MUD is designed to estimate users' signals, i.e., the expected photon counts.

The inherent nonlinear characteristics of Poisson shot noise result in a mixing of signals and noise within the received signals, making their separation and signal detection challenging. 
To address these challenges, we integrate an MUD and dedicated DECs for each user, as illustrated in Fig. \ref{fig:SystemModel}. This method diverges from frame synchronization approaches typically used under synchronous settings~\cite{OE:zhouXL2012},
as it can deal with 
varying starting points for iterative detection of frames and consequently distinct noise characteristics and external information computations. Moreover, the Poisson noise in PhC systems can introduce errors in the delay estimation. 

Given the delay estimated in Section \ref{Sec:randomaccess} and refined in Section \ref{sec:IUC-estimator}, we design the iterative MUD that operates without requiring strict frame synchronization in support of grant-free transmissions.  

\subsubsection{Estimation of PhC Noise}  
Using the interference symbols ${\Tilde{s}}_{kk',j}(l)$ derived in Section~\ref{sec:IUC-estimator}, the mean of the noise can be estimated. Let ${n}_{k,j}(l_0)$ and ${n}_{k,j}(l_1)$ represent the means of the Poisson distribution for the $l$-th slot/symbol of the $j$-th frame for user $k$ when $s_{k,j}(l) = 0$ and $s_{k,j}(l) = 1$, respectively. Specifically,  
\begin{subequations}
\label{eq:inter22c}
\begin{flalign}
    n_{k,j}(l_0) &= {n_{\mathrm{s}}} \sum_{k' \ne k} {\alpha_{k'}} G_{k'} \Tilde{s}_{kk',j}(l) + {n_{\mathrm{b}}}, \\
    n_{k,j}(l_1) &= {n_{\mathrm{s}}} G_k + {n_{\mathrm{s}}} \sum_{k' \ne k} {\alpha_{k'}} G_{k'} \Tilde{s}_{kk',j}(l) + {n_{\mathrm{b}}}.
\label{eq:inter22b}
\end{flalign}
\end{subequations}

The interference from user $k'$ in the $l$-th slot of user $k$, denoted as $\Tilde{s}_{kk',j}(l)$, is asynchronous relative to $s_{k,j}(l)$ for $k' \ne k$. This distinction is explicitly modeled in \eqref{eq:inter22c}, where synchronous scenarios can be regarded as a special case.
The total interference is expressed as
\begin{equation}
    \varsigma_{k,j}(l) \triangleq {n_{\mathrm{b}}} + {n_{\mathrm{s}}} \sum_{k' \ne k} {\alpha_{k'}}{{G}_{k'}}{{\Tilde{s}}_{kk',j}(l)}. 
\label{eq:NoiseEst1}
\end{equation}  
Using the mathematical expectation of ${\Tilde{s}}_{kk',j}(l)$, the estimated total interference becomes:
\begin{equation}
    \Tilde{\varsigma}_{k,j}(l) = {n_{\mathrm{b}}} + {n_{\mathrm{s}}} \sum_{k' \ne k} {\alpha_{k'}}{{G}_{k'}}E\left({\Tilde{s}}_{kk',j}(l)\right). 
\label{eq:NoiseEst}
\end{equation}  

By definition, we have
\begin{equation}
\label{eq:mean}
E({\tilde s_{kk',j}}(l)) = \Pr ({\tilde s_{kk',j}}(l) = 1) \cdot 1 + \Pr ({\tilde s_{kk',j}}(l) = 0) \cdot 0,  
\end{equation}  
where ${{\Tilde{s}}_{kk',j}(l)}$ is the expected number of photons under the Poisson distribution. Since $\Pr({{\Tilde{s}}_{kk',j}(l)}=1) + \Pr({{\Tilde{s}}_{kk',j}(l)}=0) = 1$, the \textit{a-priori} LLR~\cite[Eq.~(5)]{OE:zhouXL2012} is defined as:
\begin{equation}
    a\left(\Tilde{s}_{kk',j}(l)\right) \triangleq \ln \frac{\Pr(\Tilde{s}_{kk',j}(l)=1)}{\Pr(\Tilde{s}_{kk',j}(l)=0)}.
\label{eq:Le11}
\end{equation}
By substituting \eqref{eq:Le11} into \eqref{eq:mean}, it follows that 
\begin{equation}
    E\left({{\Tilde{s}}_{kk',j}(l)} = 1\right) = \frac{\exp\left[a\left(\Tilde{s}_{kk',j}(l)\right)\right]}{1 + \exp\left[a\left(\Tilde{s}_{kk',j}(l)\right)\right]}.
\label{eq:px1}
\end{equation}

\subsubsection{External Iterative Information Detection}

To address PhC noise \(\Tilde{\varsigma}_{k,j}(l)\), the extrinsic LLR is estimated using the maximum \textit{a-posteriori} probability principle. As outlined in~\cite[Sec.~3]{OE:zhouXL2012}, the iterative MUD generates extrinsic LLRs to enhance decoding in the DECs, and the DECs send extrinsic LLRs to refine the MUD. 

Let \(a\left( {{{\mathbf{s}}_k}} \right)\) and \(e\left( {{{\mathbf{s}}_k}} \right)\) denote the \textit{a-priori} and extrinsic LLRs for the MUD, respectively. 
Since the input symbols are independent,
the extrinsic LLR of the MUD is~\cite[Eq.(5)]{OE:zhouXL2012}
\begin{equation}\label{eq:MUD_post}
    e\left( s_{k,j}(l) \right) = {{\alpha} _{k}(j)}\ln \frac{\Pr\left( r_{k,j}(l) \mid s_{k,j}(l) = 1 \right)}{\Pr\left( r_{k,j}(l) \mid s_{k,j}(l) = 0 \right)}.
\end{equation}

From~\cite[Eq.(6)]{OE:zhouXL2012}, the updated extrinsic LLR becomes
\begin{equation}\label{eq:LeMud}
    e\left( {s_{k,j}(l)} \right) \!\!=\!{{\alpha}_{k}(j)}\! \big[r_{k,j}(l)\ln \frac{{n} _{k,j}(l_1)}{{n} _{k,j}(l_0)} \!\!+\!\left( {n} _{k,j}(l_0)\!\!-\!{n} _{k,j}(l_1) \right)\big].
\end{equation}
By substituting \eqref{eq:inter22c} and \eqref{eq:NoiseEst} into \eqref{eq:LeMud}, the extrinsic LLR \(e\left( s_{k,j}(l) \right)\) can be expressed as 
\begin{subequations}
\begin{align}
    e\left( s_{k,j}(l) \right) 
    \!\!=&\! {{\alpha}_{k}(j)}\!\Big[\! r_{k,j}(l)\!\ln\!\! \frac{{n_{\mathrm{s}}} G_k \!\! +\! \!{n_{\mathrm{s}}}\! \sum_{k' \!\ne k}\!{\alpha_{k'}} G_{k'}\! \Tilde{s}_{kk',j}(l) \!\!+ \!\!{n_{\mathrm{b}}}}{{n_{\mathrm{s}}} \sum_{k' \ne k}{\alpha_{k'}} G_{k'} \Tilde{s}_{kk',j}(l) \!\!+\! {n_{\mathrm{b}}}} \nonumber\\
    &- \!n_{\mathrm{s}} G_k \Big] \label{eq:Le_A} \\
    =& {{\alpha}_{k}(j)}\! \left[ r_{k,j}(l)\ln \frac{{n_{\mathrm{s}}} G_k + {\Tilde{\varsigma}_{k,j}(l)}}{{\Tilde{\varsigma}_{k,j}(l)}} - n_{\mathrm{s}} G_k \right] \label{eq:Le_B} \\
    =& {{\alpha}_{k}(j)}\! \left[ r_{k,j}(l)\ln \left( 1 + \frac{{{{n_{\mathrm{s}}}}{G_k}}}{{\Tilde{\varsigma}_{k,j}(l)}} \right) - {{n_{\mathrm{s}}}}{G_k} \right], \label{eq:Le}
\end{align}
\end{subequations}
where~\eqref{eq:Le_A} is derived by substituting \eqref{eq:inter22c} into \eqref{eq:LeMud},~\eqref{eq:Le_B} is obtained by applying \eqref{eq:NoiseEst} to \eqref{eq:Le_A},~\eqref{eq:Le} simplifies \eqref{eq:Le_B} using logarithmic operations, and~\eqref{eq:Le} represents the soft logarithmic value corresponding to the received photon count.

\subsection{Detection Algorithm and Complexity Analysis}

    \begin{algorithm}[t]
\caption{Iterative MUD for grant-free PhC OWC}
\begin{algorithmic}[1]
\State \textbf{Initialization:}
\Statex \quad Set maximum iterations $T_{\mathrm{out}}$;
\Statex \quad Input parameters: $\mathbf{r}_{k,j}$, $\mathbf{p}_m$, $\mathbf{q}$, $n_{\mathrm{b}}$, $n_{\mathrm{s}}$, $\forall k$;
\Statex \quad Prior information: $\hat{\alpha}_{k}(j-1)$, $\hat{\alpha}_{k}(j)$, $\hat{\alpha}_{k}(j+1)$, 
$\hat{\xi}_{k}(j-1)$, $\hat{\xi}_{k}(j)$, $\hat{\xi}_{k}(j+1)$,
and LLRs $\left\{ a(\mathbf{s}_{k,j}) \right\}, \forall k$;
\For{$t_{\mathrm{out}} = 1$ to $T_{\mathrm{out}}$}
    \State Compute $\left\{ \Tilde{s}_{{kk'}, j} \right\}, \forall k,k'$ using \eqref{eq:interA}--\eqref{eq:interB};
    \State Estimate noise $\left\{ \Tilde{\varsigma}_{k,j} \right\}, \forall k$ using \eqref{eq:NoiseEst}--\eqref{eq:px1};
    \State Generate extrinsic LLRs, $\left\{ e(\mathbf{s}_{k,j}) \right\}, \forall k$ by \eqref{eq:Le};
    \State Deinterleave $\left\{ e(\mathbf{s}_{k,j}) \right\}, \forall k$ to obtain $\left\{ a(\mathbf{c}_{k,j}) \right\}, \forall k$;
    \State Perform decoding to compute $\left\{ e(\mathbf{c}_{k,j}) \right\}, \forall k$, see~\cite{OE:zhouXL2012}; 
    \State Interleave $\left\{ e(\mathbf{c}_{k,j}) \right\}$ to update $\left\{ a(\mathbf{s}_{k,j}) \right\}, \forall k$;
    
\EndFor
\State \textbf{Output:} Recovered bit sequence $\left\{ \hat{\mathbf{d}}_{k,j} \right\}, \forall k$.
\end{algorithmic}
\label{algo:ParallelIterative}
\end{algorithm}

Based on the detection window and iterative MUD described in Sections~\ref{sec:IUC-estimator} and \ref{sec:AMUI-Detector}, the iterative MUD algorithm for the grant-free, multi-user, underwater PhC OWC system is summarized in \textbf{Algorithm~\ref{algo:ParallelIterative}}.
In the algorithm, Step~4 performs interference estimation using interfering symbols identified from the detection window. In Step 5, the MUD discriminates users and generates extrinsic LLRs. In Step 6, the DECs conduct soft decoding to produce extrinsic LLRs, which are fed back to the MUD to refine its user discrimination accuracy. After iteratively executing these steps, the recovered bit sequence is obtained through a hard decision in the DEC.

The computational complexity of \textbf{Algorithm \ref{algo:ParallelIterative}} arises primarily from two operations:
\begin{itemize}
    \item \textit{Interfering User Symbol Identification (Step 3):} The detection window slides for $L_{\mathrm{s}}$ slots for each of the $(K_{\mathrm{a}}-1)$ potential interfering users~\cite{TWC:jiangSC2022,hoang:2017TCOM}. This repeats for $K_{\mathrm{a}}$ active users, incurring a complexity of $\mathcal{O}\big(L_{\mathrm{s}}(K_{\mathrm{a}}-1)K_{\mathrm{a}}\big)$.
    \item \textit{Asynchronous MUI Estimation (Step 4):} For each of the $L_{\mathrm{s}}$ slots, a summation of $(K_{\mathrm{a}}-1)$ elements is performed for each of the $K_{\mathrm{a}}$ active users. Thus, the overall complexity is  $\mathcal{O}\big(L_{\mathrm{s}}(K_{\mathrm{a}}-1)K_{\mathrm{a}}\big)$.
\end{itemize}
These operations are iteratively executed for $T_{\mathrm{out}}$ iterations. Consequently, the total computational complexity of \textbf{Algorithm \ref{algo:ParallelIterative}} is 
$\mathcal{O}\big(T_{\mathrm{out}}L_{\mathrm{s}}(K_{\mathrm{a}}-1)K_{\mathrm{a}}\big)=\mathcal{O}\big(T_{\mathrm{out}}L_{\mathrm{s}}K_{\mathrm{a}}^2\big)$.
    
    \subsection{Iterative Receiver Convergence Analysis}
    Mutual information (MI) provides an accurate and robust criterion for analyzing iterative convergence~\cite{TCOM:2001Brink} and has been widely adopted. The external information transfer (EXIT) diagram visualizes the information exchange between component decoders during iterations, reflecting the feedback mechanism of extrinsic information. Prior studies have shown that EXIT analysis effectively characterizes iterative decoding convergence and predicts BER.

    For each user, the MUD and DEC modules in Fig.~2 are modeled as two independent soft-input soft-output (SISO) components. The MI between the transmitted symbol $s \in \{0,1\}$ and the \textit{a-priori} LLR $L_{\mathrm{A}}$ is given by~\cite{OE:2013performance} 
    \begin{equation}
    \begin{split}
        & I_{\mathrm{A}} = \frac12 \sum_{s \in \{0,1\}} \int_{-\infty}^{+\infty} \Pr(L_{\mathrm{A}}|S=s) \cdot \\
        & \ \ \ \ \ \ \ \ \ \log_2 \frac{2 \Pr(L_{\mathrm{A}}|S=s)}{\Pr(L_{\mathrm{A}}|S=0) + \Pr(L_{\mathrm{A}}|S=1)}\,dL_{\mathrm{A}} . \label{eq:Ia_a}
    \end{split}
    \end{equation}
    The MI between the transmitted symbol $s \in \{0,1\}$ and the extrinsic LLR $L_{\mathrm{E}}$ is given by~\cite{OE:2013performance} 
    \begin{equation}
    \begin{split}
        & I_{\mathrm{E}} = \frac12 \sum_{s \in \{0,1\}} \int_{-\infty}^{+\infty} \Pr(L_{\mathrm{E}}|S=s) \cdot \\
        & \ \ \ \ \ \ \ \ \ \log_2 \frac{2 \Pr(L_{\mathrm{E}}|S=s)}{\Pr(L_{\mathrm{E}}|S=0) + \Pr(L_{\mathrm{E}}|S=1)}\,dL_{\mathrm{E}} .\label{eq:Ie_a}
    \end{split}
    \end{equation}
    The conditional PDFs, $\Pr(L_{\mathrm{A}}|S=s)$ and $\Pr(L_{\mathrm{E}}|S=s)$, are obtained via Monte Carlo histogram statistics; the integrals can be numerically approximated without assuming specific distributions.

    \begin{figure}[t]
        \centering
        \includegraphics[width=0.8\columnwidth]{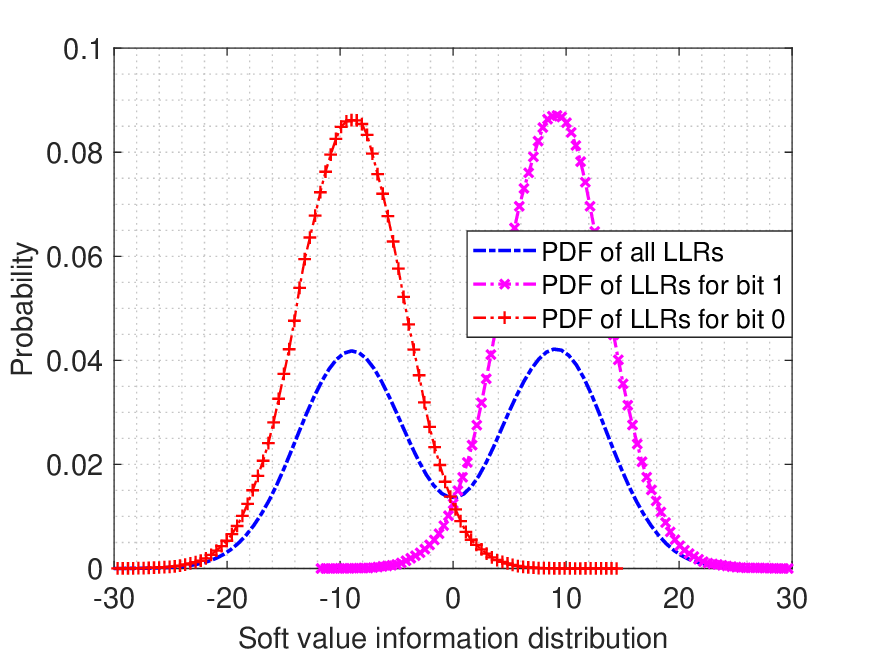}
        \caption{LLRs soft value information distribution.}
        \label{Fig_LLRs}
    \end{figure}

    \begin{figure}[t]
        \centering
        \includegraphics[width=0.90\columnwidth]{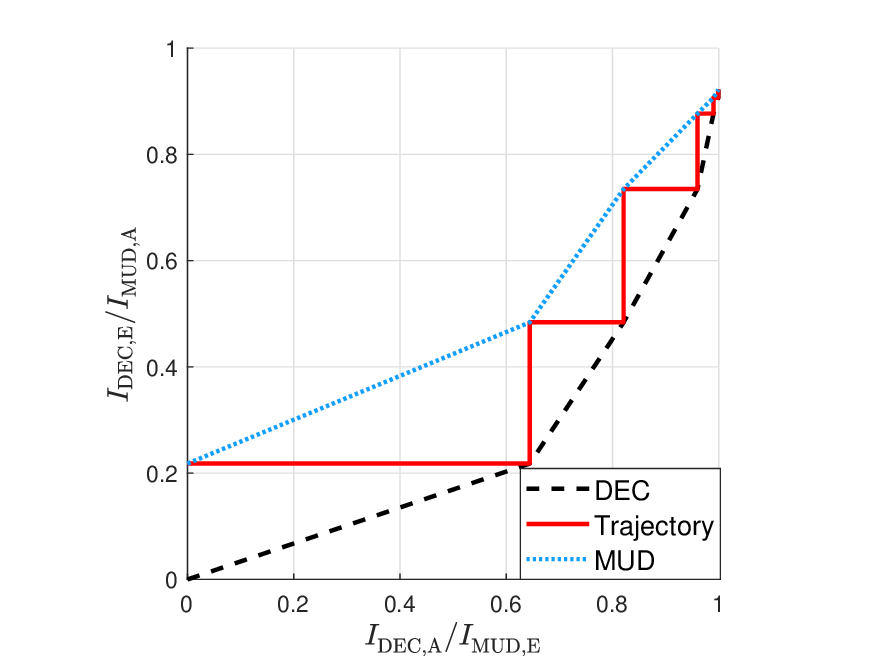}
        \caption{EXIT charts of the proposed asynchronous schemes.}
        \label{Fig_EXIT}
    \end{figure}

    Under signal-dependent Poisson shot noise, the \textit{a-priori} and extrinsic LLRs, $L_{\mathrm{A}}$ and $L_{\mathrm{E}}$, remain nearly statistically independent and their distributions are approximately Gaussian~\cite{wiberg1996codes}. As shown numerically in Fig.~\ref{Fig_LLRs}, the red and magenta curves represent the empirical PDFs of $L_{\mathrm{A}}$, which closely resemble a Gaussian PDF. Thus, $L_{\mathrm{A}}$ can be modeled as  
    \begin{equation}
        {L_{\rm{A}}} = {\mu _{\rm{A}}} \cdot (2S -1)+ {n_{\rm{A}}}, \label{eq:La_a}
    \end{equation}
    where $n_{\mathrm{A}}$ is Gaussian with variance $\sigma_{\mathrm{A}}^2$ and $\mu_{\mathrm{A}}=\sigma_{\mathrm{A}}^2/2$. Also, the symmetry and consistency conditions hold~\cite{TIT:2002Design}:
    \begin{subequations}
    \label{eq:PrLa}
    \begin{flalign}
        \Pr \left( { - {L_{\rm{A}}}\left| {S = 1} \right.} \right) & = \Pr \left( {{L_{\rm{A}}}\left| {S = 0} \right.} \right), \\
        \Pr \left( {{L_{\rm{A}}}\left| {S = s} \right.} \right) & = \Pr \left( { - {L_{\rm{A}}}\left| {S = s} \right.} \right) \cdot {e^{(2s-1) \cdot {L_{\rm{A}}}}}.
    \label{eq:Prlab}
    \end{flalign}
    \end{subequations}
    Substituting \eqref{eq:PrLa} into \eqref{eq:Ia_a}, the MI, ${{I}_{\mathrm{A}}}$, relies only on ${{\sigma }_{\mathrm{A}}}$, i.e.,
    \begin{equation}
        J\left( \sigma  \right) = {I_{\rm{A}}}\left( {{\sigma _{\rm{A}}} = \sigma } \right), \label{eq:J}
    \end{equation}    
    where $J\left( \sigma \right)$ satisfies $\underset{s\to 0}{\mathop{\lim }}\,J\left( \sigma \right)=0, \underset{s\to \infty }{\mathop{\lim }}\,J\left( \sigma \right)=1, \sigma >0$. $J\left( \sigma \right)$ is a monotonically increasing function with respect to the parameter $\sigma $, and its inverse function is
    \begin{equation}
        {\sigma _{\rm{A}}} = {J^{ - 1}}\left( {{I_{\rm{A}}}} \right). \label{eq:mu_b}
    \end{equation}

    In the iterative process, the extrinsic output of the MUD is deinterleaved and fed to the DEC as \textit{a-priori} information, and vice versa. Therefore, at the $i$-th iteration, we have
    \begin{equation}  
    \begin{cases} 
        I_{{\mathrm{DEC}},A}^{(i)} = I_{\mathrm{MUD,E}}^{(i)}, \\ 
        I_{\mathrm{MUD,A}}^{(i)} = I_{\mathrm{DEC,E}}^{(i - 1)}. 
    \end{cases} \label{eq:mutual}
    \end{equation}

    The iterative trajectory generated from \eqref{eq:mutual} constitutes the EXIT graph. The intersection of the two curves indicates the convergence point. The MI at this point predicts the BER: the closer the abscissa is to 1, the lower the BER. If the intersection occurs at the MI of 1, error-free recovery is theoretically achievable. As shown in Fig.~\ref{Fig_EXIT}, the EXIT trajectory converges rapidly within five iterations, confirming convergence. This is further corroborated by the BER simulations in Fig.~\ref{fig:Iter}, where the BER performance converges within five iterations.


   \section{Numerical Results}
    \label{sec:rsults}

    \begin{table}[t]
    \caption{Simulation parameter configuration }
    \label{table:SIMULATION PARAMETERS}
        \centering
        \begin{tabular}{ >{\centering\arraybackslash}m{0.5cm} 
        >{\raggedright\arraybackslash}m{5cm} >{\centering\arraybackslash}m{2cm}}
         \hline
           & Definition & Value \\ 
         \hline
         $K$ & Number of users & 10 \\ 
         $M$ & Number of groups & 2 \\
         $\tau $ & Slot duration & 1 $\mu {\mathrm{s}}$ \\
         $N_{\mathrm{c}}$ & Length of spread spectrum & 10\\
         ${L_{\mathrm{b}}}$ & Length of data & 1024 bits\\ 
         ${L_{\mathrm{p}}}$ & Length of SP after spreading & 511 bits\\ 
         ${L_{\mathrm{q}}}$ & Length of VP after spreading & 320 bits\\
         ${\alpha _k}$ &  Activity of user $k$ & 0.5 \\        
         ${\varepsilon _{\mathrm{p}}}$ & Delay estimation threshold & 0.75\\ 
         ${\varepsilon _{\mathrm{q}}}$ & Delay verification threshold & 0.3\\
         $G_k$ & Channel gain & ${I_k} {L_k}$\\
         $h$ & Planck's constant & $6.626 \times 10^{34} \ {\mathrm{J}} \cdot {\mathrm{s}}$ \\ 
         $\eta$ & Quantum efficiency & 0.5 \\ 
         ${f}$ & Optical center frequency & 600 THz\\
         $T_{\mathrm{in}}$ & Maximum number of verification rounds & 4\\
         $T_{\mathrm{out}}$ &  Maximum number of receiver iterations & 12\\
        \hline
        \end{tabular}
    \end{table}

    In this section, we use Monte-Carlo simulations to evaluate the proposed synchronization algorithm (Algorithm \ref{algo:BayesDelayEst}) and detection algorithm (Algorithm \ref{algo:ParallelIterative}). 
    The simulation was conducted at the system level using MATLAB to model the complete underwater OWC system. The BS is positioned at the bottom of the underwater platform.
    The system includes 10 users randomly distributed along a water pipe, with water depth ranging from 5 m to 45 m~\cite{TCOM:ChenYK2023}. The underwater communication channel is characterized by a clean seawater environment~\cite{GLOBECOM2011channel}, where background radiation depends on water depth and quality, and is assumed to be -165 dBJ~\cite{TCOM:ChenYK2023}.

    \begin{figure*}[htp]
		\centering
		\subfigure[]{
			\centering
			\includegraphics[width=0.31\textwidth]{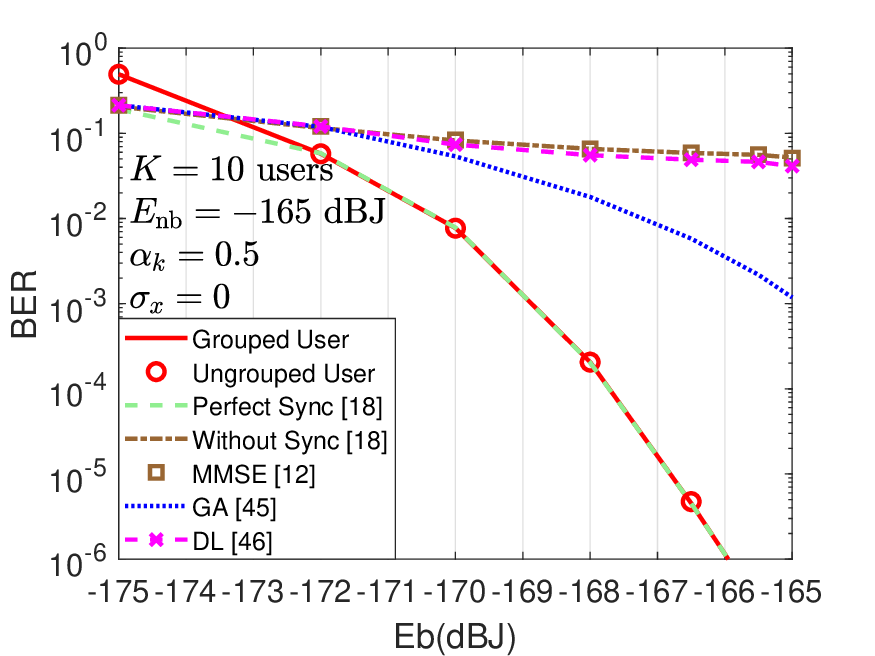}
			\label{BER_0fading}
		}
		\subfigure[]{
			\centering
			\includegraphics[width=0.31\textwidth]{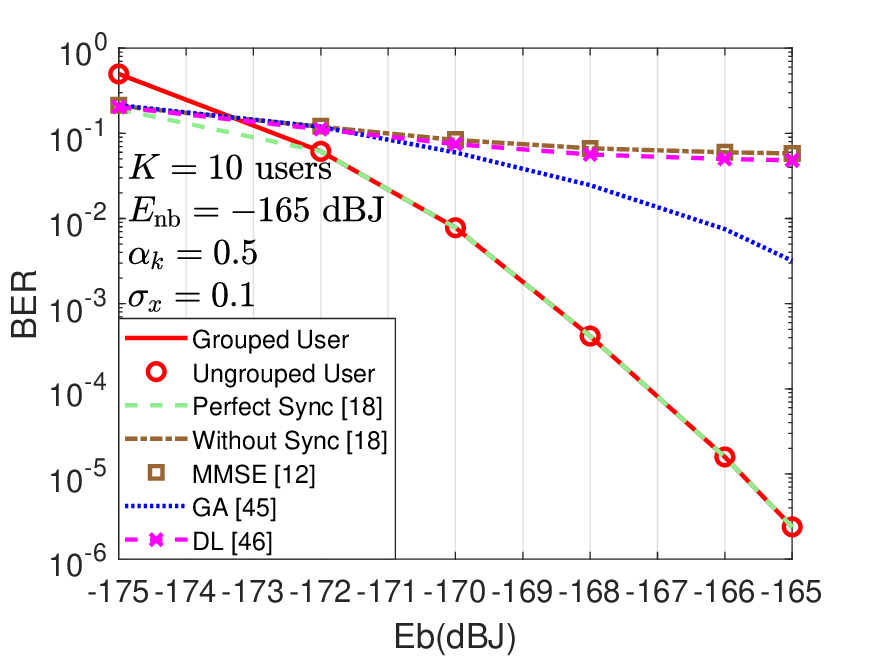}
			\label{BER_0.1fading}
        }
        \subfigure[]{
			\centering
			\includegraphics[width=0.31\textwidth]{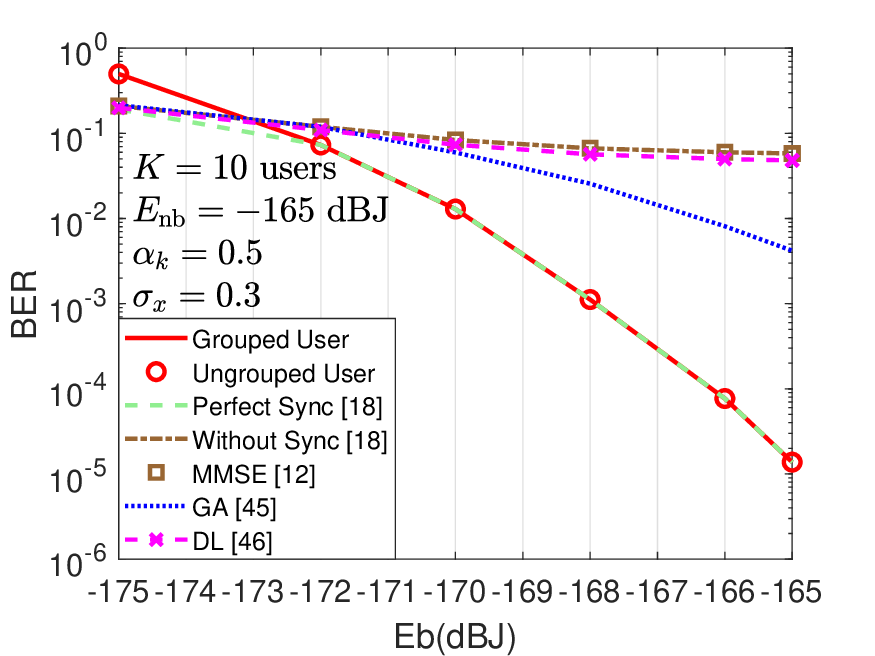}
			\label{BER_0.3fading}
		}
		\caption{BER comparison curves of the two proposed asynchronous schemes and the benchmark schemes during the communication process. (a)  fading intensity $\sigma_{x}=0$. (b) fading intensity $\sigma_{x}=0.1$. (c) fading intensity $\sigma_{x}=0.3$.}
    \label{fig:Ber}
	\end{figure*}

We set the slot duration to $\tau = 1\ \mu s$ and choose a blue-green laser with an optical center frequency of 600 THz~\cite{COMST:2021UnderwaterSurvey}, selected for its low attenuation coefficient in seawater~\cite{JOC:2013monte}. The signal energy received per slot is ${E_{\mathrm{b}}} \triangleq {P_{\mathrm{s}}}\tau$, which is the total energy from all $K$ users' lasers hitting a single PD. The background radiation energy per slot, denoted as ${E_{\mathrm{nb}}}$, incorporates environmental factors and the PD's dark current. It is linearly proportional to the expected number of background photons, ${n_{\mathrm{b}}}$, as given by ${E_{\mathrm{nb}}} \triangleq {n_{\mathrm{b}}}{hf}$. The spread spectrum length, data length, SP length after spreading, and VP length after spreading of each frame are set to ${N_{\mathrm{c}}} = 10$, ${L_{\mathrm{b}}} = 1024 \text{ bits}$, ${L_{\mathrm{p}}} = 511 \text{ bits}$, and ${L_{\mathrm{q}}} = 320 \text{ bits}$, respectively. Consequently, the frame length after spreading is 11.081 ms. The user activity probability is assumed to be $\alpha = 0.5$. Simulation parameters are summarized in Table~\ref{table:SIMULATION PARAMETERS}; $10^4$ independent Monte-Carlo tests are performed per simulation.


To the best of our knowledge, none addresses asynchronous multi-user detection in photon-counting Poisson channels with the proposed Bayesian delay estimation and verification framework. In particular, the existing Gaussian approximation (GA) and minimum mean square error (MMSE) methods assume independent noise and signals. However, in the Poisson noise model, they are inseparable, making these methods ineffective for signal detection. We consider the following benchmarks all under the assumption that user activity is known:
\begin{itemize}
    \item 
    \textit{Iterative MUD With Perfect Synchronization (Perfect Sync):} In this benchmark, the delays of all users are perfectly known to the BS, and applied to the iterative detection algorithm extended from~\cite{OE:zhouXL2012}. 
    
    \item \textit{Iterative MUD Without Synchronization (Without Sync):} 
    Adapted from the method developed in~\cite{OE:zhouXL2012}, this algorithm is an iterative detection method without delay estimation even if the signals are received asynchronously at the BS. It helps validate the effectiveness of the proposed delay estimation and synchronization algorithms.
    
    \item \textit{MMSE Without Synchronization:} This is a classical MUD method extended by assuming an approximate SNR under the Poisson model, i.e., $n_{\mathrm{s}}/\sqrt{n_{\mathrm{s}}/2 + n_{\mathrm{b}}}$~\cite{TCOM:ChenYK2023}. This method optimizes the signal detection by minimizing the mean square error between the received and estimated signals. However, the method does not account for delay estimation in the asynchronous scenario. 
    
    \item \textit{Gaussian Approximation With Synchronization:} Assuming that the delays of all users are known, this benchmark estimates the mean and variance of inter-user interference by treating it as Gaussian noise, instead of evaluating each user's contribution. This is a classic method for MUD under the Gaussian assumption~\cite{WCN:Liping2003}.    

    \item \textit{Deep Learning (DL) Without Synchronization:} Adapted from a DL-based detection algorithm developed in~\cite{TWC:2025Uzlaner} under a Gaussian noise model, this benchmark integrates a deep neural network with traditional receiver processing and uses fixed model parameters for detection without delay estimation or synchronization. This benchmark helps evaluate the proposed asynchronous receiver.

\end{itemize}
All users share a common SP in conventional delay estimation. To assess the group-level delay estimator developed in Section~\ref{sec:GL-estimator}, we compare the delay estimation between both grouped and ungrouped users.

Fig. \ref{fig:Ber} shows the BER performance of the proposed 
iterative MUD method, i.e., Algorithm 2, 
compared to the benchmarks.
It is observed that, as \( E_{\mathrm{b}} \) increases, the proposed algorithm with both grouped and ungrouped user designs achieves the desired BER comparable to the iterative MUD with perfect synchronization.
This improvement is attributed to two key factors: the efficient correlation operations for group-level and user-level delay estimation in Algorithm \ref{algo:BayesDelayEst}, which provide accurate user delay estimates, and the precise identification of interfering symbols in Algorithm \ref{algo:ParallelIterative}, which reduces interference during iterative MUD. 
Although the BERs of the proposed algorithm with both grouped and ungrouped user designs are slightly lower than those of the iterative MUD without synchronization, Gaussian approximation, MMSE, and DL without synchronization algorithms at very low \( E_{\mathrm{b}} \), 
the advantage of the proposed algorithm becomes apparent as \( E_{\mathrm{b}} \) increases. This is because inter-user interference becomes the dominant noise source, negatively affecting the performance of the Gaussian approximation, MMSE, iterative MUD without synchronization, and DL without synchronization. 

    \begin{figure}[t]
        \centering
        \includegraphics[width=0.77\columnwidth]{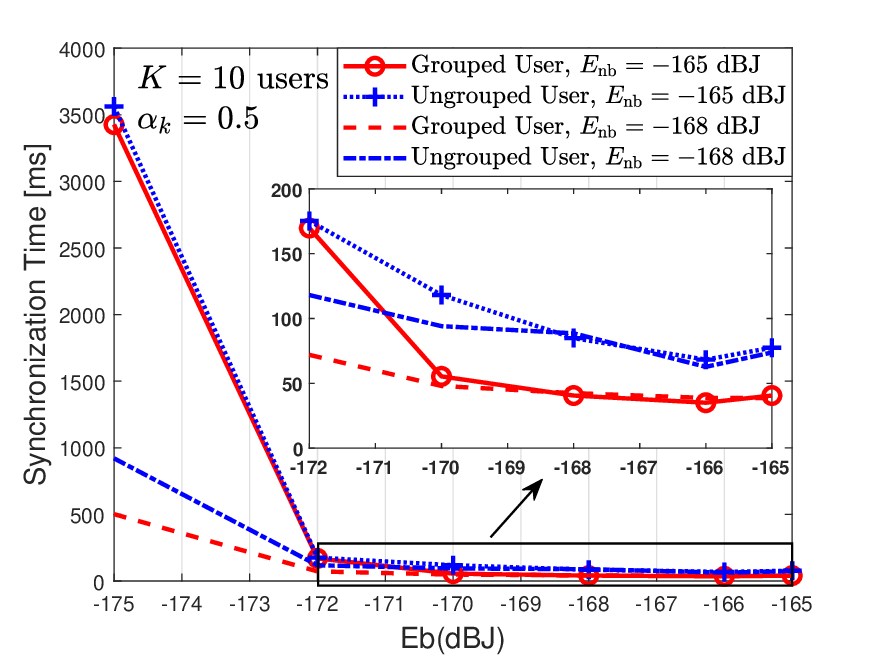}
        \caption{Synchronization time of the two proposed asynchronous schemes during the communication establishment process.}
        \label{fig:time}
    \end{figure}

    \begin{figure}[t]
        \centering
        \includegraphics[width=0.77\columnwidth]{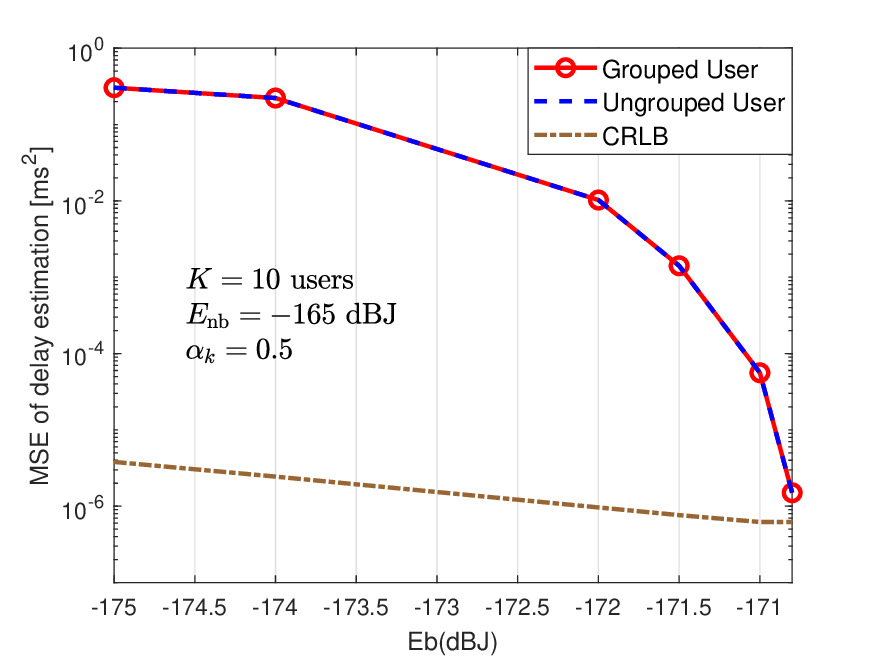}
        \caption{\color{blue}{Delay estimation MSE comparision curves of the two proposed asynchronous schemes and CRLB.}}
        \label{fig:Mse_Eb}
    \end{figure}

    Fig. \ref{fig:time} shows the synchronization time of the proposed scheme with grouped and ungrouped user designs. The synchronization time is defined as the time required for all users to synchronize and detect the SP. Our scheme quickly establishes communication for all 10 users. At very low \( E_{\mathrm{b}} \), the synchronization time increases for both grouped and ungrouped user designs. However, as \( E_{\mathrm{b}} \) increases, the synchronization time decreases significantly, with communication established within 5 frames at the fastest rate. This improvement is due to the ability of {Algorithm \ref{algo:BayesDelayEst}} to accurately estimate user delays and activity, as \( E_{\mathrm{b}} \) increases. Moreover, the synchronization time of our scheme with grouped user design is halved compared to the ungrouped user design, with the fastest establishment time of shorter than 50 ms. This efficiency arises because the group-level delay estimation in {Algorithm~\ref{algo:BayesDelayEst}} reduces the number of users in each group, minimizing conflict probability and enhancing synchronization accuracy and speed.  

Comparing Figs. \ref{fig:Ber} and \ref{fig:time}, we observe that the low-complexity scheme with grouped users (i.e., the grouped user scenario) in {Algorithm \ref{algo:BayesDelayEst}} achieves a shorter synchronization time than its counterpart with ungrouped users, with minimal performance loss. This further highlights the effectiveness of the user grouping strategy proposed in Section III-A.

    \begin{figure}[t]
        \centering
        \includegraphics[width=0.77\columnwidth]{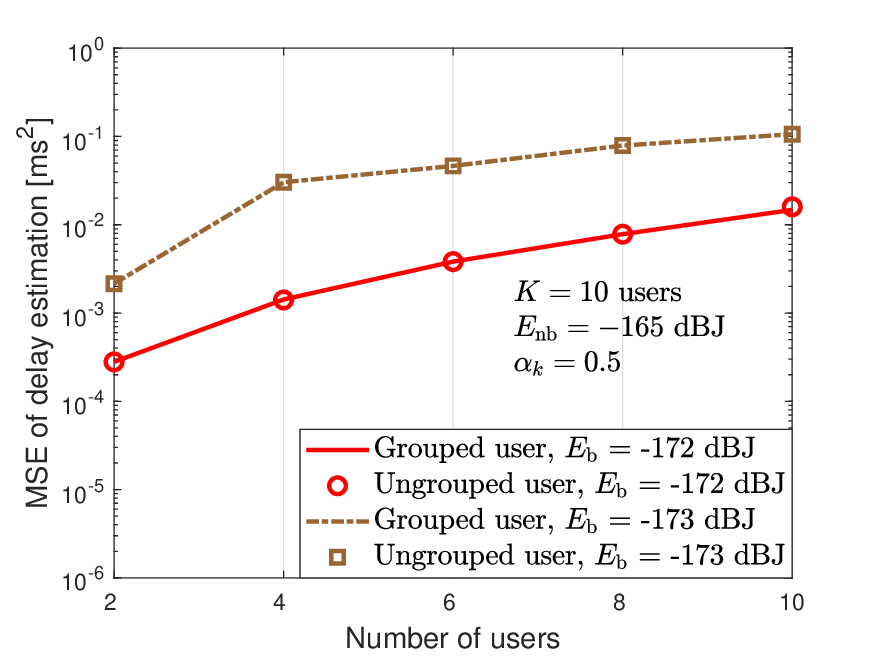}
        \caption{Delay estimation MSE curves of the two proposed asynchronous schemes under different numbers of users.}
        \label{fig:Mse_ms}
    \end{figure}

    \begin{figure}[t]
        \centering
        \includegraphics[width=0.77\columnwidth]{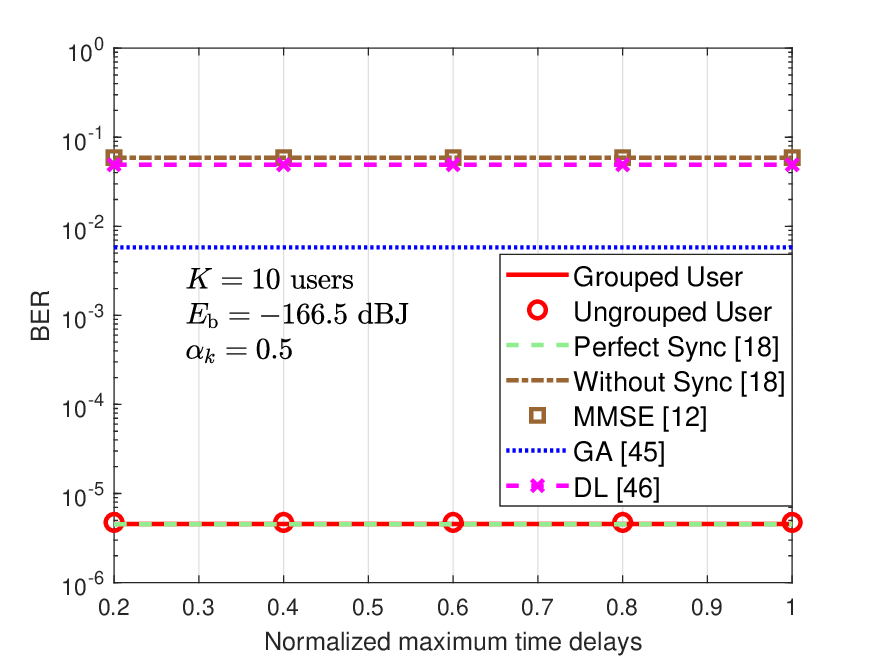}
        \caption{BER curves of the two proposed asynchronous schemes and the benchmark schemes under different maximum time delays of the users.}
        \label{fig:delay}
    \end{figure}

    \begin{figure}[t]
        \centering
        \includegraphics[width=0.77\columnwidth]{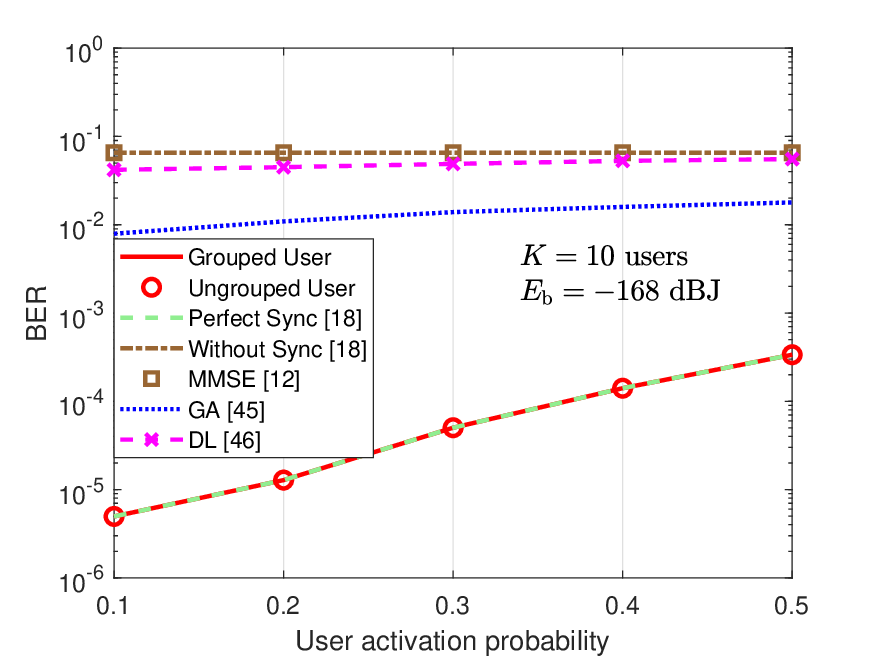}
        \caption{BER curves of the two proposed asynchronous schemes and the benchmark schemes different user activation probabilities.}
        \label{Fig:activity}
    \end{figure}

    \begin{figure}[t]
        \centering
        \includegraphics[width=0.77\columnwidth]{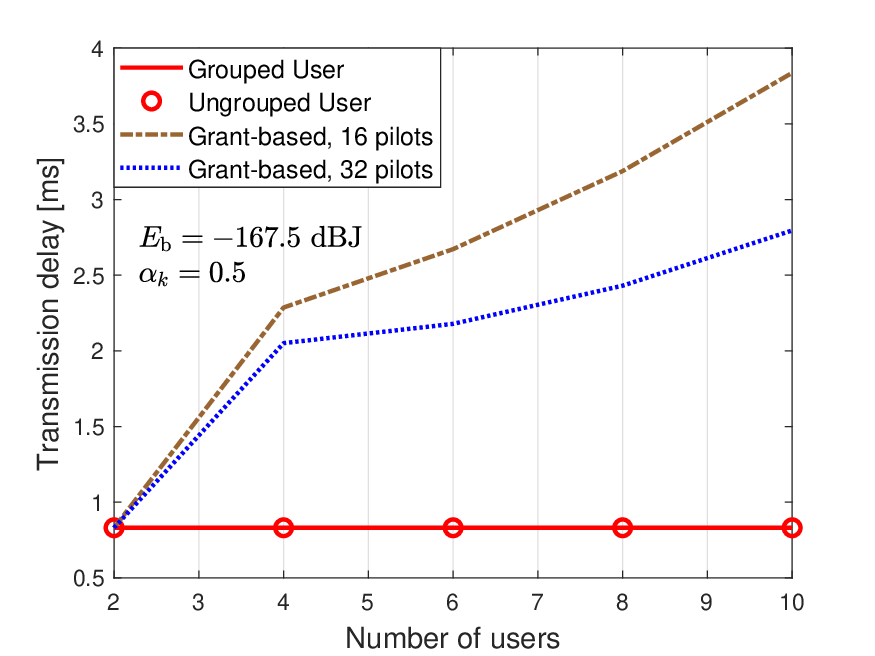}
        \caption{Transmission delays of the proposed grant-free scheme and the grant-based benchmark under different numbers of users.}
        \label{Fig:Transmission_delay}
    \end{figure}

    \begin{figure}[t]
        \centering
        \includegraphics[width=0.77\columnwidth]{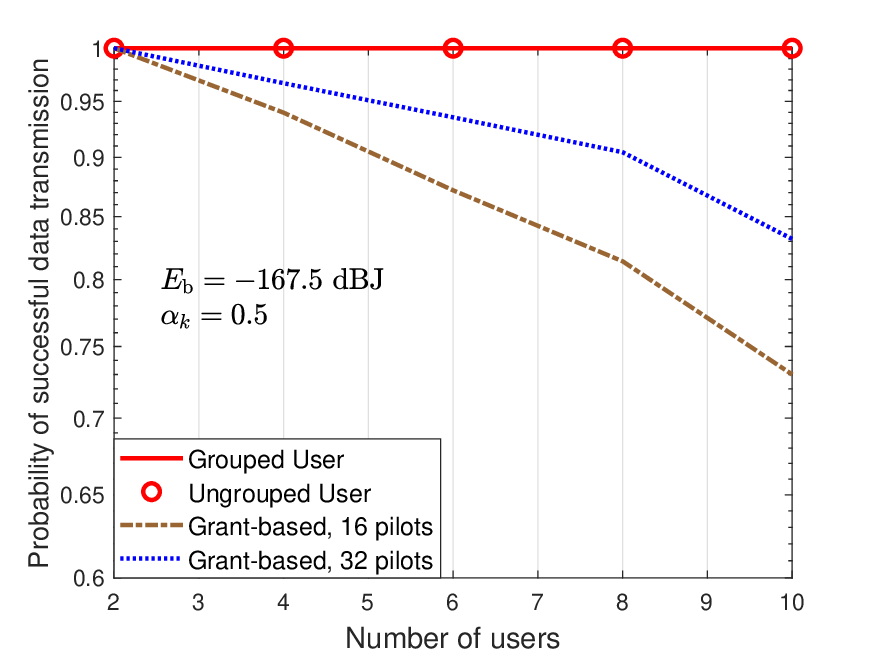}
        \caption{Data transmission success rates of the proposed grant-free scheme and the grant-based benchmark under different numbers of users.}
        \label{Fig:collision}
    \end{figure}

    \begin{figure}[t]
        \centering
        \includegraphics[width=0.77\columnwidth]{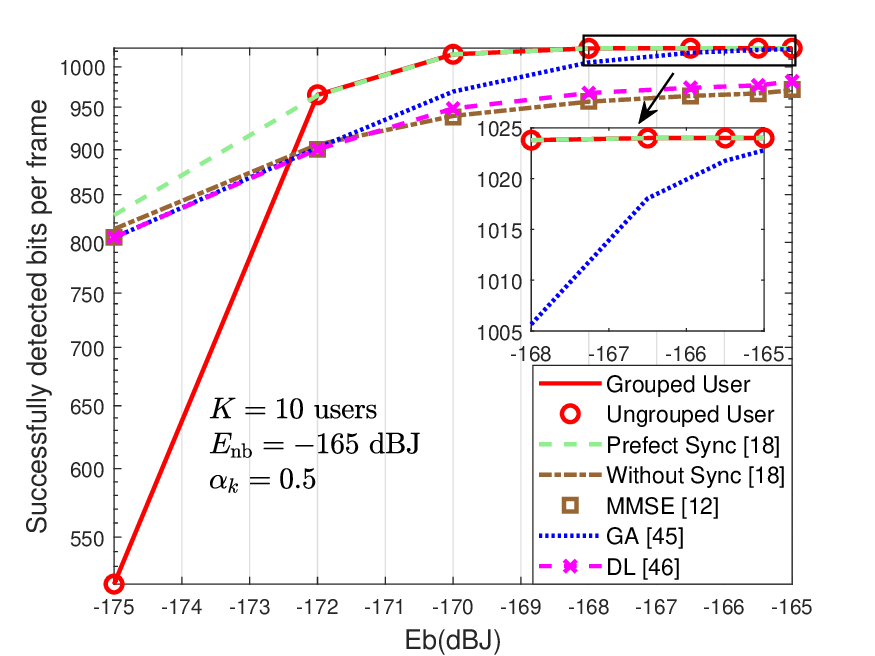}
        \caption{Comparison in the number of successfully detected bits per frame between the two proposed asynchronous schemes and the benchmarks.}
        \label{Fig:success_bits}
    \end{figure}
    
    \begin{figure}[t]
        \centering
        \includegraphics[width=0.77\columnwidth]{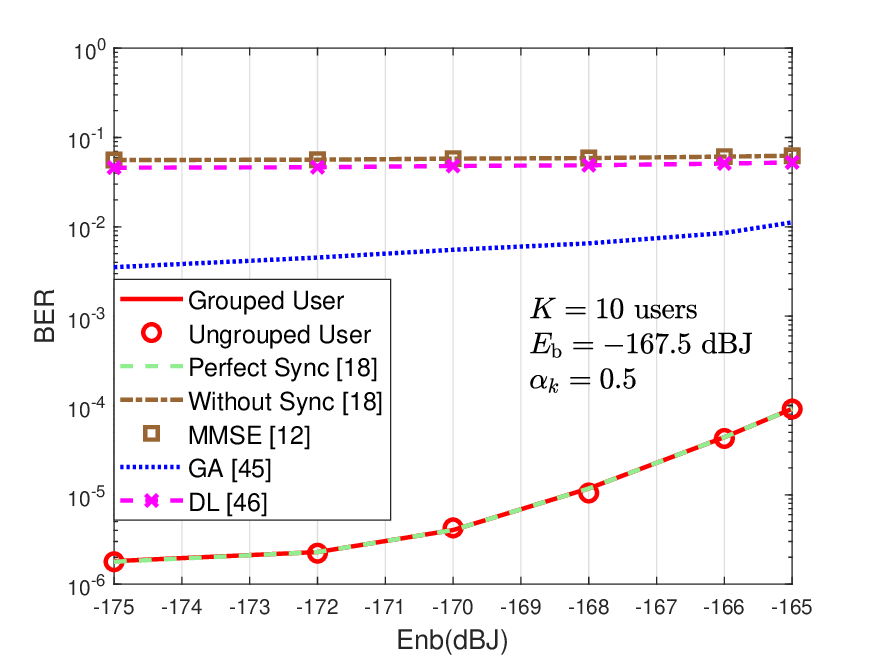}
        \caption{ BER comparison curves of the two proposed asynchronous schemes and the benchmark schemes under different background radiations.}
        \label{fig:nb}
    \end{figure}

    \begin{figure}[t]
        \centering
        \includegraphics[width=0.77\columnwidth]{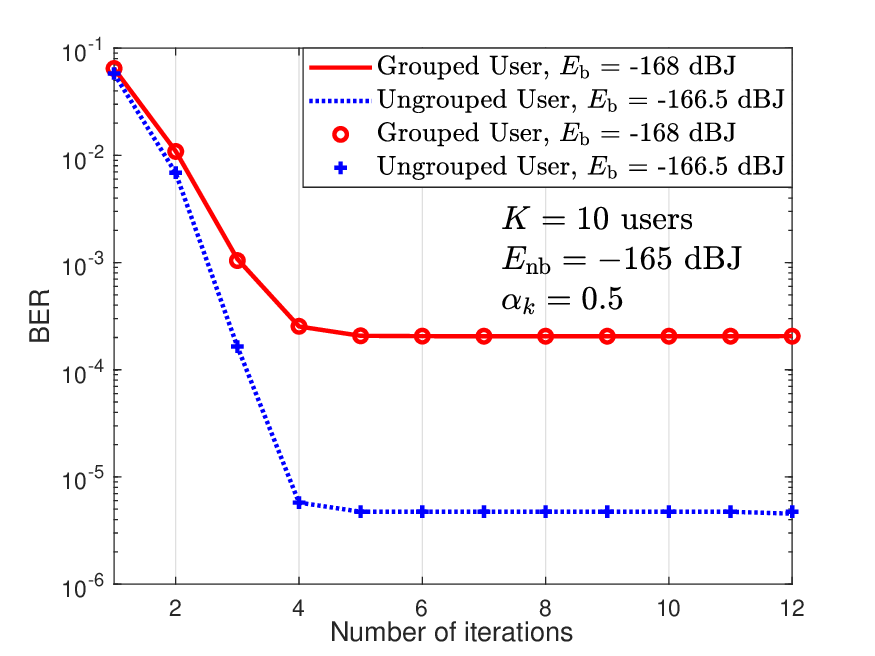}
        \caption{Convergence comparison curves of the two proposed asynchronous schemes during the communication process.}
        \label{fig:Iter}
    \end{figure}

    \begin{figure}[t]
        \centering
        \includegraphics[width=0.77\columnwidth]{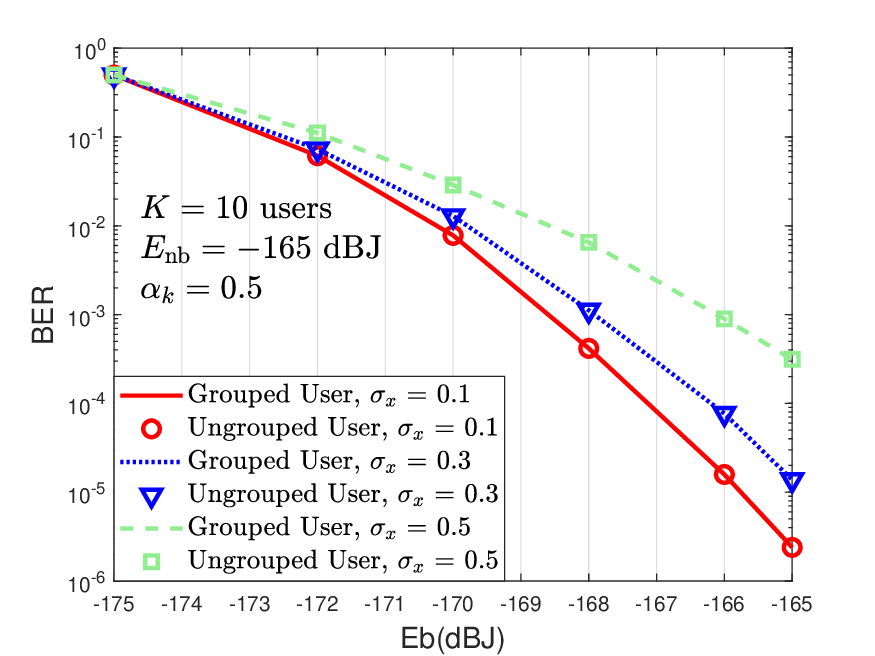}
        \caption{BER comparison curves of the two proposed asynchronous schemes under different turbulence intensities.}
        \label{fig:fading}
    \end{figure}

    \begin{figure}[t]
        \centering
        \includegraphics[width=0.77\columnwidth]{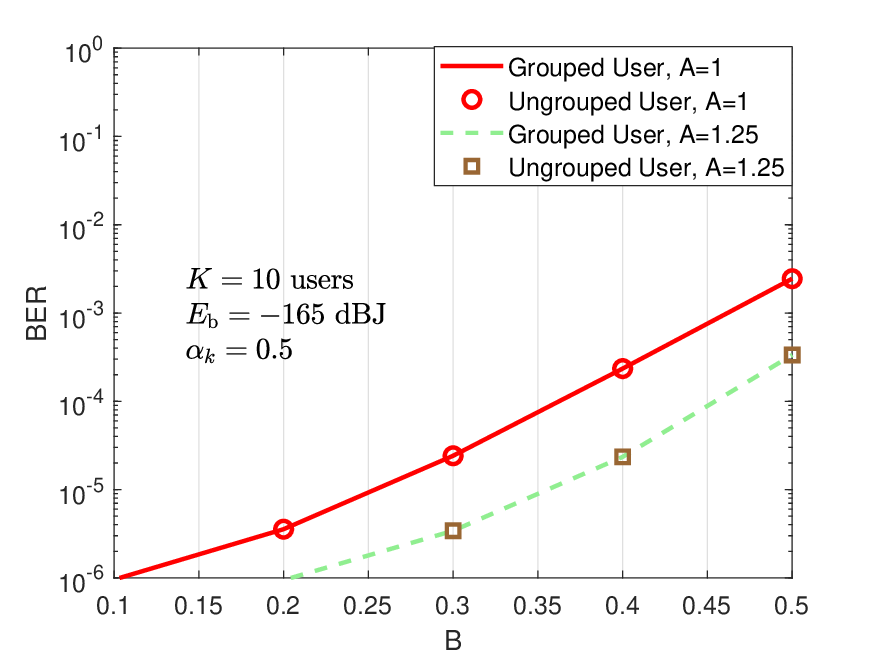}
        \caption{\color{blue}The performance loss of the two proposed asynchronous schemes under imperfect CSI.}
        \label{Fig_CSI}
    \end{figure}

    {\color{blue}Fig.~\ref{fig:Mse_Eb} shows the mean squared error (MSE) of the proposed scheme under asynchronous transmission with the CRLB plotted for comparison. As $E_{\mathrm{b}}$ increases, the MSE performance of the proposed algorithm with both grouped and ungrouped user designs approaches that of CRLB. This improvement is primarily attributed to the efficient correlation operations of the delay estimation algorithm's group-level and user-level delay estimation and delay check, which effectively suppress multi-user interference and Poisson shot noise.
    Although the MSE of the proposed algorithm with both grouped and ungrouped user designs is higher than that of CRLB at very low $E_{\mathrm{b}}$, the difference is negligible as communication becomes infeasible under these conditions, as shown in Fig. 8. As $E_{\mathrm{b}}$ increases, the advantages of the proposed algorithm become increasingly prominent.}
    
    Fig.~\ref{fig:Mse_ms} shows the delay estimation MSE performance of the proposed schemes. Under extremely low energy conditions ($E_{\mathrm{b}} \le -173$ dBJ), the MSE of both the grouped and ungrouped user designs increases with the number of users. However, as shown in Fig. \ref{fig:Ber}, communication is no longer feasible within this energy range, so the impact on the system is negligible. As $E_{\mathrm{b}}$ increases, the MSE decreases rapidly, indicating that the group-level and user-level delay estimation and delay verification designed in Algorithm 1 can achieve accurate user delay estimation.

    Fig.~\ref{fig:delay} shows the BER performance versus maximum user delays, simulated over the normalized range $[0.2, 1] L_{\mathrm{s}}$ (equivalent to frame length). The proposed algorithm with grouped and ungrouped user designs maintains robust performance across a wide range of delays, achieving a BER close to that under perfect synchronization. This demonstrates that the delay estimation method and frame structure can effectively mitigate the impact of asynchronous transmission delays.

    Fig. \ref{Fig:activity} shows the BER performance of the two proposed asynchronous schemes compared to the benchmarks under different user activation probabilities. While the BER slightly increases with higher activation probability due to increased multi-user interference, the proposed algorithm with both grouped and ungrouped user designs achieves the desired BER comparable to the iterative MUD with perfect synchronization. Notably, even with an activation probability of $\alpha_k = 0.5$, the proposed algorithm achieves a BER below $10^{-3}$ at \( E_{\mathrm{b}} = -168\) dBJ. This demonstrates the robustness of the proposed algorithm in handling high user activity, which is attributed to the accurate delay estimation that effectively mitigates MUI.

    Fig.~\ref{Fig:Transmission_delay} shows the transmission delay of the proposed grant-free scheme and the grant-based benchmark under different numbers of users. The delay of the proposed grant-free scheme mainly comes from the overhead of SP and VP pilots, while the delay of the grant-based scheme includes the overhead of request, grant, queuing and collision retransmission. As the number of users increases, the benchmark grant-based NOMA system shows a significant delay increase in both 16-pilot and 32-pilot configurations. By contrast, the proposed scheme always maintains a low fixed overhead. This is attributed to the fact that the grant-free mechanism eliminates the request-grant process, avoiding additional transmission delay.

    Fig.~\ref{Fig:collision} compares the data transmission success rates of the proposed grant-free scheme and the grant-based benchmark under different user numbers. The proposed grant-free scheme achieve 100\% success rates across all user scales, while the grant-based benchmark scheme success rate decreases with the increasing number of users due to collisions and retransmissions. This demonstrates the advantage of the proposed grant-free design in mitigating collisions. 

    Fig.~\ref{Fig:success_bits} shows the average number of successfully detected bits per frame for the proposed iterative MUD scheme and the benchmarks under different received powers. As \( E_{\mathrm{b}} \) increases, both the proposed grouped and ungrouped user designs can recover all transmitted bits, achieving performance comparable to the perfect synchronization scheme. Although the number of successfully detected bits is slightly lower than that of the perfect synchronization scheme at very low \( E_{\mathrm{b}} \), the advantage of the proposed algorithm becomes apparent as \( E_{\mathrm{b}} \) increases. This is consistent with the BER analysis in Fig.~8. Thanks to the accurate user delay estimation of Algorithm 1 and the precise identification of interference symbols by Algorithm 2, the interference during the iterative MUD period is reduced. 

Fig. \ref{fig:nb} plots the BER performance of the different schemes under different background radiation levels. The proposed scheme with either grouped or ungrouped user design performs comparably to the iterative MUD scheme with perfect synchronization.
As the background radiation,  \( {E_{\mathrm{nb}}} \), increases, the BER degrades due to the increased invalid counts, such as background radiation and dark counts. However, our scheme maintains a significant performance advantage over the Gaussian approximation and iterative MUD without synchronization. Specifically, under various background radiation conditions, our scheme accurately estimates user delays and activity, while distinguishing interfering symbols. In accordance with~\cite{TCOM:ChenYK2023}, the system sustains excellent BER performance even under low \( E_{\mathrm{nb}} \) conditions at a receiver depth of 10 m (corresponding to \( {E_{\mathrm{nb}}} = -165 \ \mathrm{dBJ} \)).

Fig. \ref{fig:Iter} shows the convergence behaviors of the considered schemes under different \( E_{\mathrm{b}} \) values. The proposed scheme with either grouped or ungrouped user design converges within up to five iterations. As \( E_{\mathrm{b}} \) increases, convergence slows down due to the higher photon counter saturation, which supports more photons for transmission and exacerbates inter-user interference. As a consequence, more iterations are required.

Fig. \ref{fig:fading} plots the BERs of the proposed scheme under varying turbulence intensities. As the turbulence intensity \( \sigma_k \) increases from 0 to 0.5, the BERs of the scheme degrade. Nonetheless, even at \( \sigma_k = 0.5 \), the scheme maintains a desirable BER of lower than \( 10^{-3} \) when \( E_{\mathrm{b}} = -165 \ \mathrm{dBJ} \), demonstrating that the proposed scheme exhibits a good level of robustness against turbulent fading. 

        {\color{blue}We further evaluate the system's robustness under imperfect CSI. Let ${\varepsilon_k}$ denote the CSI error of user $k$, modeled as a Gaussian random variable with zero mean and variance $\psi_k^2$. Following a classical CSI error model~\cite{TCOM:2015aquilina}, $\psi_k = \left( B G_k \right)^{A}$, where $A$ and $B$ account for the error level. The estimated CSI is $ \hat{G}_k = G_k + \varepsilon_k$. Perfect CSI corresponds to $B = 0$ or $A \to \infty$, i.e., $\varepsilon_k = 0$.
        Fig.~\ref{Fig_CSI} plots the BER performance of the proposed algorithm under imperfect CSI. When $A = 1$, the BER performance of the proposed grouped and ungrouped user designs degrades slightly as $B$ increases. Nevertheless, the proposed algorithm maintains a BER below $10^{-3}$ under moderate CSI errors (e.g., $A = 1.25$, and $B \leq 0.5$). The robustness of the proposed single-antenna scheme is demonstrated to realistic CSI imperfections in underwater communication scenarios.
        }

	\section{Conclusion}
    \label{sec:conclusion}
    In this paper, we have proposed a new algorithm for delay estimation and signal recovery from received signals in a grant-free, multi-user, underwater PhC OWC system with signal-dependent Poisson shot noise. Our synchronization algorithm employs a specialized frame structure to accurately estimate user delay while considering signal-dependent Poisson shot noise, and then feeds these delay estimates into our iterative MUD which identifies interfering symbols, estimates MUI, and conducts signal detection on a slot basis. Extensive simulations have validated that our scheme achieves BER performance on par with ideal iterative detection algorithms.

    \bibliographystyle{IEEEtran}
    \bibliography{IEEEabrv,reference}

\end{document}